\documentclass[reprint,superscriptaddress,amsmath,amssymb,aps]{revtex4-2}
\usepackage{graphicx}% Include figure files
\usepackage{dcolumn}% Align table columns on decimal point
\usepackage{bm}% bold math
\usepackage{dcolumn}

\usepackage{color}

\def\cbl{\color{blue}}

\definecolor{darkblue}{rgb}{0,0.02,0.45}
\definecolor{darkred}{rgb}{0.45,0.02,0} 

\usepackage[colorlinks,bookmarks=false,citecolor=blue,linkcolor=red,urlcolor=blue]{hyperref}
\usepackage{multirow}

\def\be{\begin{equation}}
\def\ee{\end{equation}}
\def\bea{\begin{eqnarray}}
\def\eea{\end{eqnarray}}

\newcommand{\ra}[1]{\renewcommand{\arraystretch}{#1}}

\def\bs{\boldsymbol}
\def\mc{\mathcal}

\def\jsaf{J_\text{s}^{\text{AF}}}
\def\jsfm{J_\text{s}^{\text{FM}}}
\def\jwaf{J_\text{w}^{\text{AF}}}
\def\jwfm{J_\text{w}^{\text{FM}}}
\def\jooaf{J_\text{O$\cdots$\!{O}}^{\text{AF}}}

\usepackage{verbatim}

\let\maketitlesup\maketitle
\begin{document}
\title{Quintuplet condensation in the skyrmionic insulator Cu$_2$OSeO$_3$\\ at ultrahigh magnetic fields}

\author{T. Nomura}
\email{nomura.toshihiro@shizuoka.ac.jp}
\affiliation{Department of Physics, Shizuoka University, Shizuoka 422-8529, Japan}
\author{I. Rousochatzakis}
\email{I.Rousochatzakis@lboro.ac.uk}
\affiliation{Department of Physics, Loughborough University, Loughborough LE11 3TU, United Kingdom}
\author{O. Janson}
\email{o.janson@ifw-dresden.de}
\affiliation{Leibniz Institute for Solid State and Materials Research Dresden, 01609 Dresden, Germany}
\author{M. Gen}
\affiliation{Institute for Solid State Physics, University of Tokyo, Kashiwa, Chiba 277-8581, Japan}
\author{X.-G.~Zhou}
\affiliation{Institute for Solid State Physics, University of Tokyo, Kashiwa, Chiba 277-8581, Japan}
\author{Y. Ishii}
\affiliation{Institute for Solid State Physics, University of Tokyo, Kashiwa, Chiba 277-8581, Japan}
\author{S. Seki} 
\affiliation{Research Center for Advanced Science and Technology, University of
Tokyo, Tokyo 113-8656, Japan}
\affiliation{Department of Applied Physics, The University of Tokyo, Tokyo 113-8656, Japan}
\author{Y. Kohama}
\affiliation{Institute for Solid State Physics, University of Tokyo, Kashiwa, Chiba 277-8581, Japan}
\author{Y. H. Matsuda}
\affiliation{Institute for Solid State Physics, University of Tokyo, Kashiwa, Chiba 277-8581, Japan}

\date{\today}

\begin{abstract}
We report ultrahigh magnetic field Faraday rotation results on the chiral helimagnet Cu$_2$OSeO$_3$, the first Mott insulator showing skyrmion lattice phases and a linear magnetoelectric effect. Between 180 and 300\,T, we find signatures of a Bose-Einstein condensation (BEC) of magnons, which can be described as a canted XY ferrimagnet. Due to the magnetoelectric coupling, the transverse magnetic order of the individual Cu$^{2+}$ spins is accompanied by a characteristic dome-like electric polarization which is crucial for the observation of the condensate via the Faraday rotation effect.
\end{abstract}

\maketitle

{\it Introduction.} 
The chiral helimagnet Cu$_2$OSeO$_3$ is one of the most celebrated materials hosting magnetic skyrmions in an insulating medium~\cite{PMID:22499941}.
The space group of Cu$_2$OSeO$_3$ is $P2_13$, in common with the prototypical skyrmion hosting systems (e.g., MnSi and Fe$_{1-x}$Co$_x$Si)~\cite{doi:10.1126/science.1166767,Yu2010}. The competition between the exchange and the Dzyaloshinskii–Moriya (DM) interactions stabilizes noncollinear spin textures and magnetic skyrmions at finite temperatures. 
Apart from the skyrmion phases, the magnetic field-temperature phase diagram of Cu$_2$OSeO$_3$, which has been studied extensively in the field range below 1~T~\cite{PhysRevB.85.220406,PhysRevB.85.224413,PhysRevB.95.134412,PhysRevLett.108.237204,PhysRevB.89.064406,PhysRevB.94.064418,Wu2015,PhysRevB.100.165143,Chacon2018,doi:10.1126/sciadv.aat7323,Tucker2016,Luo2020},  shows a plethora of other phases, including helical, conical, and ferrimagnetic, with the latter surviving at least up to 64 T~\cite{Ozerov2014,PhysRevB.83.052402}.

Notably, Cu$_2$OSeO$_3$ also features a linear magnetoelectric (ME) effect~\cite{PhysRevB.78.094416,PhysRevB.82.144107,PMID:22499941,PhysRevB.85.224413,PhysRevB.86.060403,Judit2014}, enabling skyrmion manipulation via electric fields without Joule heating losses~\cite{White_2012,PhysRevLett.113.107203,Okamura2016,D2NR04399H,Fert2013}. 
Furthermore, chirality of the crystal gives rise to nonreciprocal responses under magnetic fields, opening additional opportunities for device applications~\cite{PhysRevLett.114.197202,PhysRevB.93.235131,PhysRevLett.122.145901,Seki2020,doi:10.1073/pnas.2022927118}.

\begin{figure}[!t]
\centering
\includegraphics[width=0.99\columnwidth]{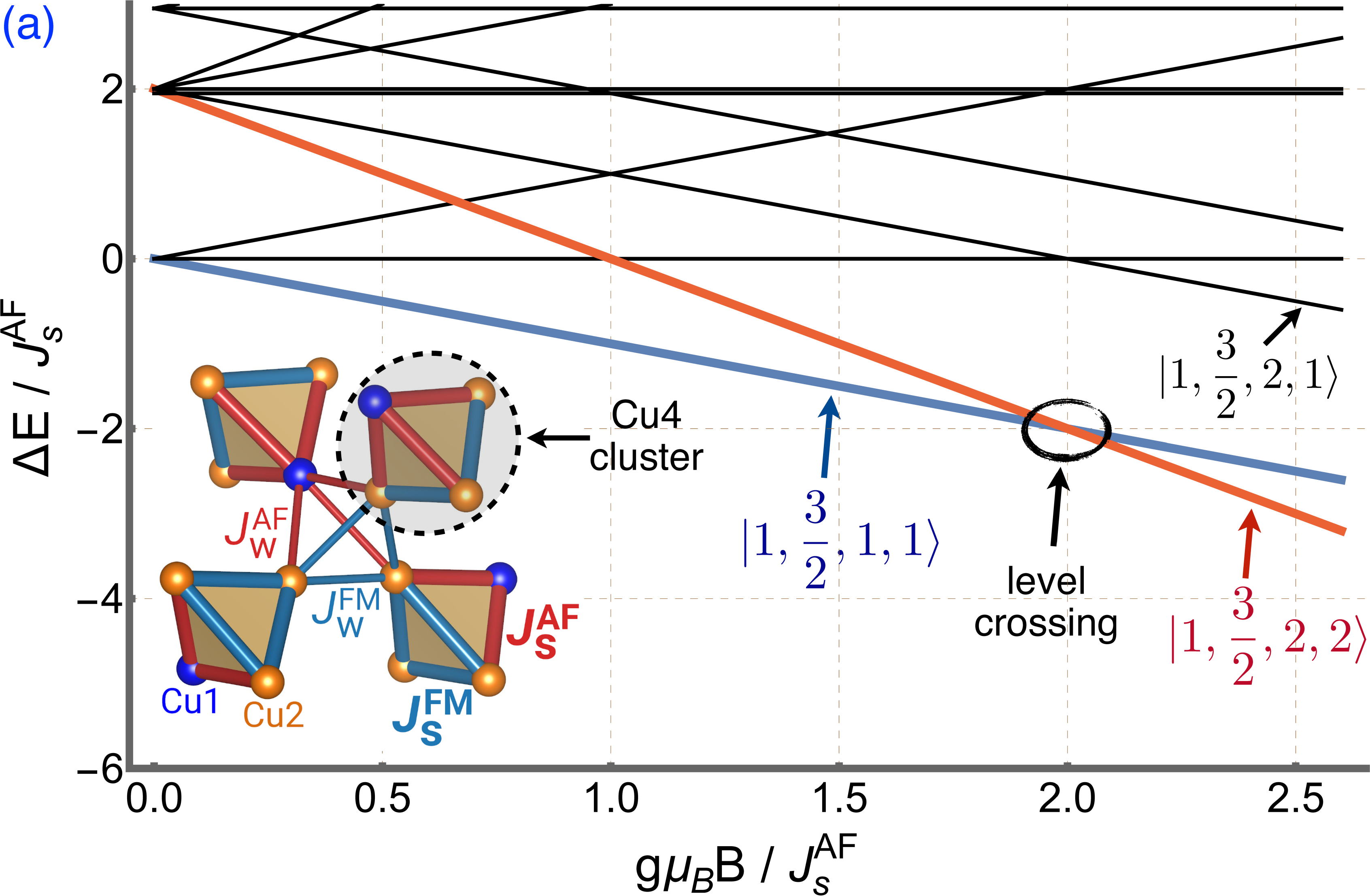}
\includegraphics[width=0.99\columnwidth]{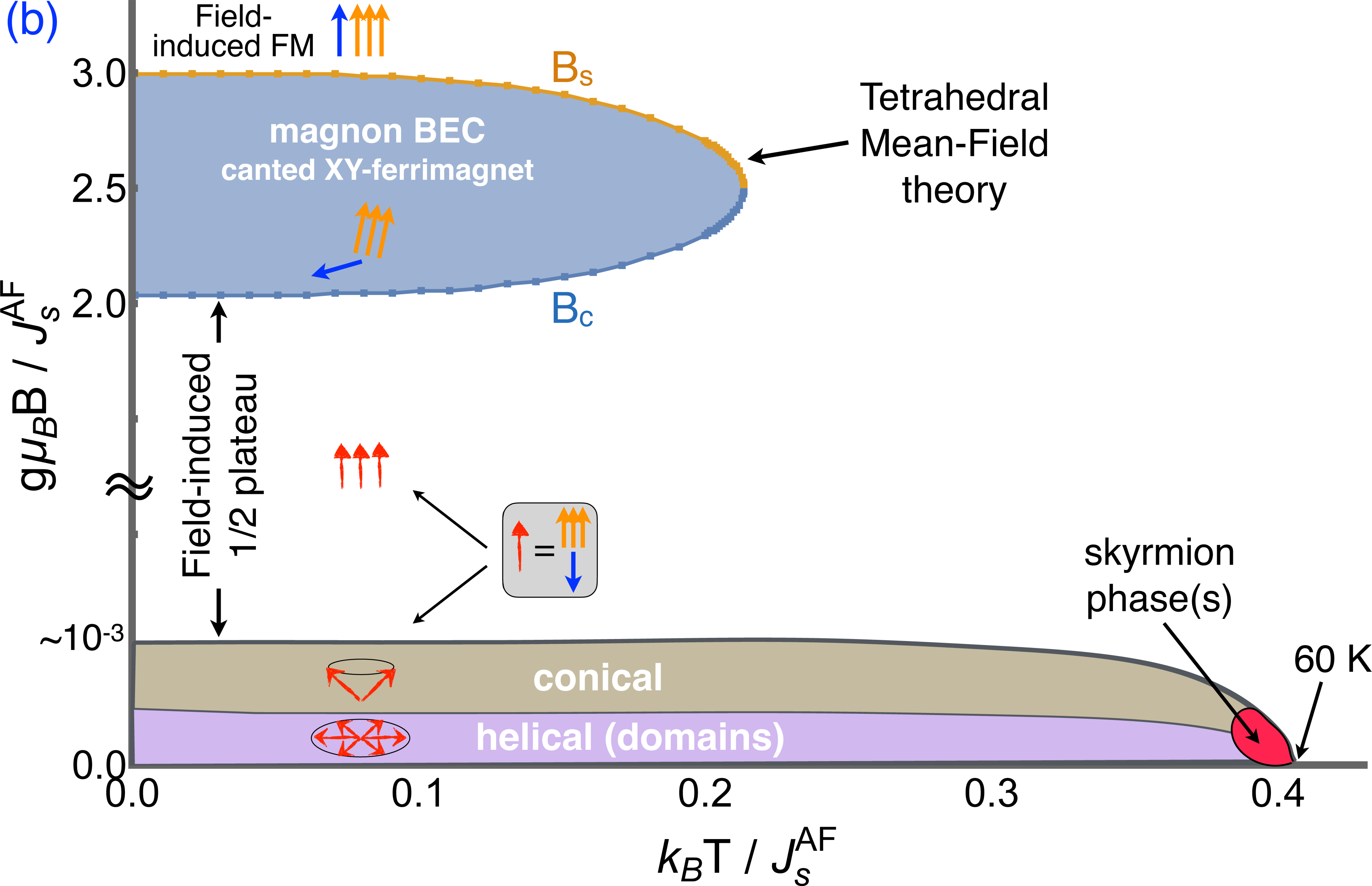}
\caption{\label{fig:H0spectrum-PD}
(a) Inset: Network of Cu$^{2+}$ ions in Cu$_2$OSeO$_3$. Cu(1) and Cu(2) represented by blue and brown spheres, respectively. The intra-tetrahedral couplings $\jsaf$ and $\jsfm$ and the inter-tetrahedral couplings $\jwaf$ and  $\jwfm$ are shown by bold and thin lines (the longer-range Cu(1)-Cu(2) coupling $\jooaf$ is not shown)~\cite{Janson2014}. The subscripts `s' and `w' stand for `strong' and `weak' couplings, respectively.
Main: Field-dependence of the energy diagram (energies measured from the zero-field ground state triplet) of an isolated Cu$_4$ cluster for the exchange parameters of Ref.~\cite{Ozerov2014}. 
(b) Temperature-field phase diagram, showing the low-field conical, helical, skyrmionic phases (schematic), and the high-field magnon condensate presented in this study (see text).
}
\end{figure}

Structurally, Cu$_2$OSeO$_3$ can be thought of as a deformed pyrochlore lattice, with one Cu(1) (Wyckoff position $4a$) and three Cu(2) ($12b$) sites comprising a single tetrahedron. As shown in Ref.~\cite{Janson2014}, and confirmed experimentally~\cite{Ozerov2014,Portnichenko2016,Tucker2016,Luo2020}, this material can be thought of as a `breathing pyrochlore' magnet (weakly coupled Cu$_4$ clusters), due to a striking separation of exchange energy scales at the microscopic level 
(Fig.~\ref{fig:H0spectrum-PD}\,(a), inset). 
This separation and, in particular, the quantum-mechanical nature of the triplet ground state of the Cu$_4$ clusters (see Fig.~\ref{fig:H0spectrum-PD}\,(a)) are indispensable for understanding a number of observed properties, from the diameter of the skyrmions and the sign of their handedness, to the strikingly low value of the ordering temperature (compared to that expected at the classical level)~\cite{Janson2014}, and the rich structure of the elementary excitation spectrum~\cite{Judit2014,Portnichenko2016,Tucker2016,Luo2020}. 
This richness contrasts with the high-field behavior, which is dominated by remarkably wide plateau at $\frac12$ of the saturation magnetization~\cite{PhysRevB.83.052402, Ozerov2014}. Further field-induced transitions, including eventual saturation of magnetization, are expected, yet they require extremely high fields to overcome the dominant antiferromagnetic exchange within Cu$_4$ clusters.

Here, we report the observation of a magnon Bose-Einstein condensate (BEC) at ultrahigh magnetic fields 180--300~T, sandwiched between the $\frac12$ magnetization plateau and the fully saturated phase (Fig.~\ref{fig:H0spectrum-PD}\,(b)). 
Such magnetic field-induced BECs have been observed in several quantum magnets (see \cite{RevModPhys.86.563,Giamarchi2008} and references therein), and are typically associated with the condensation of an excitation mode involving single ions (as in %spin-3/2 
Ba$_2$CoGe$_2$O$_7$ ($S\!=\!3/2$)~\cite{Watanabe2023} 
%spin-2 
and Ba$_2$FeSi$_2$O$_7$ ($S\!=\!2$)~\cite{PhysRevB.107.144427}) or quantum dimers (as in 
BaCuSi$_2$O$_6$ ($S\!=\!1/2$)~\cite{Jaime2004,Sebastian2006} and  TlCuCl$_3$ ($S\!=\!1/2$)~\cite{Tanaka2001,Ruegg2003,PhysRevLett.125.267207}, and 
Ba$_3$Mn$_2$O$_8$ ($S\!=\!1$)~\cite{Uchida2002,Samulon2008,Samulon2009}).  
Here, the condensation is driven by the closing of the triplet-quintuplet gap of the tetrahedra, Cu$_4$ building blocks of Cu$_2$OSeO$_3$ (Fig.~\ref{fig:H0spectrum-PD}\,(a)). 
The resulting BEC phase can be described as a canted XY ferrimagnet, in which the total transverse magnetization vanishes identically and the order parameter is associated with the transverse components of the individual Cu$^{2+}$ spins instead. Additionally, this order parameter drives a distinctive electric polarization via the linear ME coupling.

{\it Experiments.} We employed electromagnetic flux compression, generating a field of up to 500 T~\cite{Miura2003,10.1063/1.5044557}. Single crystals of Cu$_2$OSeO$_3$ were grown by the chemical vapor transport~\cite{PhysRevB.82.144107}. We polished the (110) surfaces of the D-type crystal to the thickness of $d \sim 0.6$~mm and fixed it inside the He-flow cryostat~\cite{PhysRevResearch.2.033257}. We used the 532 nm laser as a linearly polarized incident light source. We analyzed the p- and s-polarized intensities of transmitted light ($I_\mathrm{p}$ and $I_\mathrm{s}$) and obtained the Faraday rotation angle $\theta_\mathrm{F} = \arccos{(I_\mathrm{p}-I_\mathrm{s})/(I_\mathrm{p}+I_\mathrm{s})}$. Experiments were performed in the Faraday geometry with the field along the [110] direction. For details, see Supplemental Material (SM)~\cite{suppl}.

\begin{figure}[tb]
\centering
\includegraphics[width=0.99\columnwidth]{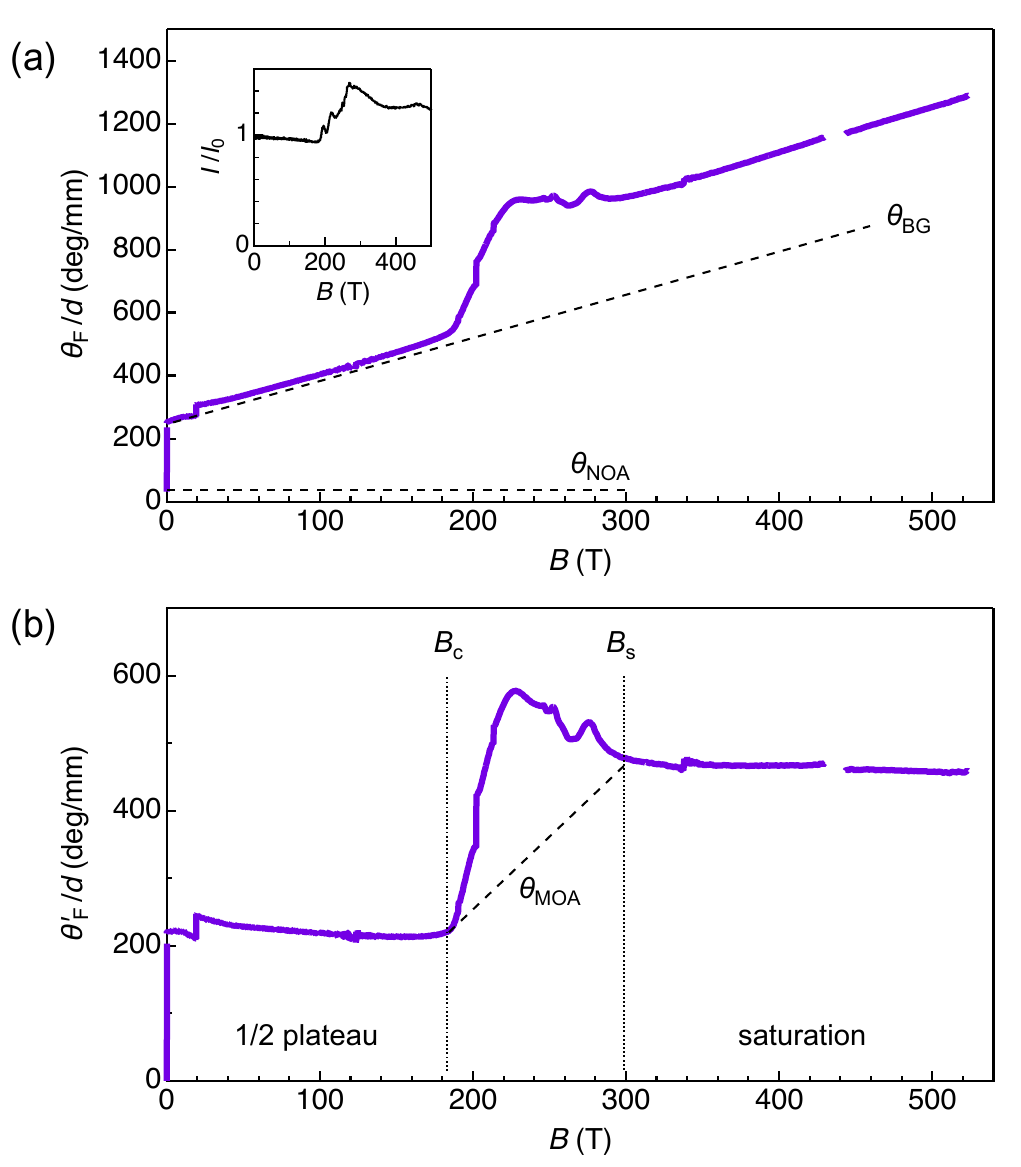}
\caption{\label{fig:fr}
Faraday rotation angle of Cu$_2$OSeO$_3$ normalized by the sample thickness, (a) $\theta_\mathrm{F}/d$ and (b) $\theta_\mathrm{F}'/d$ [Eq.~(\ref{eq:theta2})] as a function of a magnetic field at 5~K. The inset in (a) shows the transmission intensity. Contributions from the natural and magnetic optical activities ($\theta_\mathrm{NOA}$, $\theta_\mathrm{MOA}$) and the linear background ($\theta_\mathrm{BG}$) are indicated by the dashed lines. $B_\mathrm{c}$ and $B_\mathrm{s}$ indicate the end of the $\frac12$ plateau and the saturation magnetic field, respectively.
}
\end{figure}

Figure~\ref{fig:fr}\,(a) shows the Faraday rotation angle $\theta_\mathrm{F}$ of Cu$_2$OSeO$_3$ normalized by the sample thickness $d$ at 5~K up to 500~T.
The inset shows the transmitted light intensity $I=(I_\mathrm{s}^2+I_\mathrm{p}^2)^{\frac12}$ normalized by the zero-field value $I_0$ as a function of magnetic fields.
The signal intensity remains high up to 500 T, indicating that the incident linearly polarized light is not significantly disturbed and that depolarization effects, if present, are minimal under our experimental conditions.
$\theta_\mathrm{F}/d$ starts from 25 deg/mm because of the natural optical activity (NOA)~\cite{PhysRevB.94.094409}.
The rapid increase of $\theta_\mathrm{F}/d$ at 1~T reflects the magnetic optical activity (MOA) upon the helical-conical-ferrimagnetic transition~\cite{PhysRevB.94.094409}.
The linear increase of $\theta_\mathrm{F}/d$ from 1~T to 180~T is not related to the magnetization, because the magnetization does not change in this field range~\cite{PhysRevB.83.052402}.
A similar slope appears above 300~T, which is a sufficiently large magnetic field to break the antiferromagnetic couplings in Cu$_2$OSeO$_3$.
Thus, one can expect that the magnetization saturates above 300~T, and the linear increase of $\theta_\mathrm{F}/d$ exists as a background signal.
This background signal probably originates from the charge-transfer absorption edge~\cite{PhysRevB.83.052402}, whose energy can slightly shift as a function of the magnetic field and cause the additional rotation angle.

For interpreting the data, we separate the Faraday rotation components as
\begin{equation}\label{eq:theta1}
\theta_\mathrm{F} = \theta_\mathrm{NOA} + \theta_\mathrm{MOA} + \theta_\mathrm{BG} + \theta_\mathrm{D}.
\end{equation}
Here, $\theta_\mathrm{NOA}$ and $\theta_\mathrm{MOA}$ represent the natural and magnetic optical activity components, respectively.
$\theta_\mathrm{BG}$ represents the linear background component indicated by the linear slope observed at 1--180~T and 300--500~T.
$\theta_\mathrm{D}$ is a dome-like component discussed later.
For discussing the magnetization of Cu$_2$OSeO$_3$, Fig. \ref{fig:fr}\,(b) plots 
\begin{equation}\label{eq:theta2}
\theta_\mathrm{F}'/d = (\theta_\mathrm{F} - \theta_\mathrm{NOA} - \theta_\mathrm{BG})/d.
\end{equation}
Since the Faraday rotation angle reflects the magnetization, Fig. \ref{fig:fr}\,(b) indicates that the $\frac12$ plateau continues up to $B_\mathrm{c}=180(5)$~T, and the magnetization saturates at $B_\mathrm{s}=300(15)$~T.
The hump observed at 260~T is probably due to the mechanical vibration of the sample (discussed in SM~\cite{suppl}).
The maximum of $\theta_\mathrm{F}'$ observed at 220~T indicates that another component exists other than the MOA because the magnetization should not decrease with increasing $B$.

{\it Theoretical analysis.} 
We use the {\it ab initio} microscopic modeling of Refs.~\cite{PhysRevLett.109.107203,Janson2014}. For our purposes, we can safely disregard the DM anisotropy  (see below). The minimal Heisenberg model comprises five exchange couplings, $\jsaf$, $\jsfm$, $\jwaf$, $\jwfm$, and $\jooaf$ (the first four are shown in Fig.~\ref{fig:H0spectrum-PD}\,(a), inset). For the values of these couplings, we take the estimates reported in Ref.~\cite{Ozerov2014} (see, however, End Matter for a comparison between various parameter sets reported in the literature~\cite{PhysRevLett.109.107203,Janson2014,Ozerov2014,Portnichenko2016,Tucker2016}).

Our starting point is the physics of an isolated Cu$_4$ tetrahedron, described by the spin Hamiltonian
\bea\label{eq:H0}
\mc{H}_0&\!=\!&
\jsaf {\bf S}_4\!\cdot\! {\bf S}_{123} \!+\! \jsfm ({\bf S}_1\!\cdot\!{\bf S}_2\!+\!{\bf S}_2\!\cdot\!{\bf S}_3\!+\!{\bf S}_3\!\cdot\!{\bf S}_1)\!-\!h S^z\!\!\!,~~~~
\eea
where ${\bf S}_{123}={\bf S}_1+{\bf S}_2+{\bf S}_3$ is the total spin of the three Cu(2) ions,  ${\bf S}_4$ is the spin of Cu(1) ion, $S^z$ is the $z$ component of the total spin ${\bf S}={\bf S}_{123}+{\bf S}_4$, $h=g\mu_\mathrm{B} B$ (where $g\simeq2.11$~\cite{LARRANAGA20091,Belesi_2011,Ozerov2014} is the electronic $g$ factor, and $\mu_\mathrm{B}$ is the Bohr magneton),  and we have chosen the field to point along $z$. The eigenstates of $\mc{H}_0$ can be labeled by $|S_{12},S_{123},S,M\rangle\,$, where $S_{12}$, $S_{123}$, $S$, and $M$ are the quantum numbers corresponding to ${\bf S}_{12}={\bf S}_1+{\bf S}_2$, ${\bf S}_{123}$, ${\bf S}$, and $S^z$, respectively (all eigenstates are listed in End Matter).

Figure~\ref{fig:H0spectrum-PD}\,(a) shows the field dependence of the energy diagram of $\mc{H}_0$. A level crossing between the low-field ground state triplet $|1,3/2,1,1\rangle$ and the quintuplet component $|1,3/2,2,2\rangle$ is expected at $h = 2 \jsaf$, corresponding to a field of $\sim$205~T for $\jsaf=145$\,K~\cite{Ozerov2014}. This crossing causes an abrupt magnetization jump from the $\frac12$ plateau ($M=1$) to full saturation ($M=2$). 
In reality, the triplet-quintuplet excitations become mobile due to the weak inter-tetrahedral interactions ($\jwaf$, $\jwfm$, and $\jooaf$), leading to a dispersive band with a minimum at the $\Gamma$ point~\cite{Judit2014,Portnichenko2016}. As the magnetic field increases, the corresponding excitation energy is lowered and eventually becomes zero at a critical field $h_\mathrm{c}$. At this point, the ${\bf Q}=0$ triplet-quintuplet mode condenses, and long-range order in the transverse components of the spins sets in, breaking the U(1) symmetry spontaneously (Fig.~\ref{fig:H0spectrum-PD}\,(b)).

To capture this physics, we focus in the region around the level crossing between $|1,3/2,1,1\rangle$ and $|1,3/2,2,2\rangle$. Given that higher energy levels are well separated in energy, we can map this $2\times2$ manifold to that of a pseudospin $\tau=\frac12$, with 
\be\label{eq:taumapping}
|1,\frac{3}{2},2,2\rangle \mapsto |\tau^z=\frac{1}{2}\rangle\,,~
|1,\frac{3}{2},1,1\rangle \mapsto |\tau^z=-\frac{1}{2}\rangle\,.
\ee 
Similarly to analogous situations in  dimerized systems (see, e.g., \cite{MilaXXZ1998}), this mapping leads to effective interactions between the $\tau$ pseudospins that take the form of an XXZ model (here with first- and second-neighbor interactions) in a longitudinal field (see End Matter). A variational mean-field treatment of this model, in which all $\tau$ pseudospins point along the same direction, somewhere in the $xz$ plane (we pick the $x$ axis for the transverse components of the spins) reveals the presence of an intermediate phase, with critical fields $h_\mathrm{c}$ (end of $\frac12$ plateau) and $h_\mathrm{s}$ (onset of fully polarized phase) given by 
\be\label{eq:hchs}
\begin{array}{l}
h_\mathrm{c} = 2\jsaf + \frac{1}{8} ( \jwaf+\jooaf) + \frac{11}{24} \jwfm\,,\\
h_\mathrm{s} = 2(\jsaf  + \jwaf + \jooaf)\,.
\end{array}
\ee
The expression for $h_\mathrm{c}$ is approximate, whereas that of $h_\mathrm{s}$ agrees with the exact result obtained from a diagonalization in the one-magnon sector.

\begin{figure}[tb]
\centering
\includegraphics[width=0.97\columnwidth]{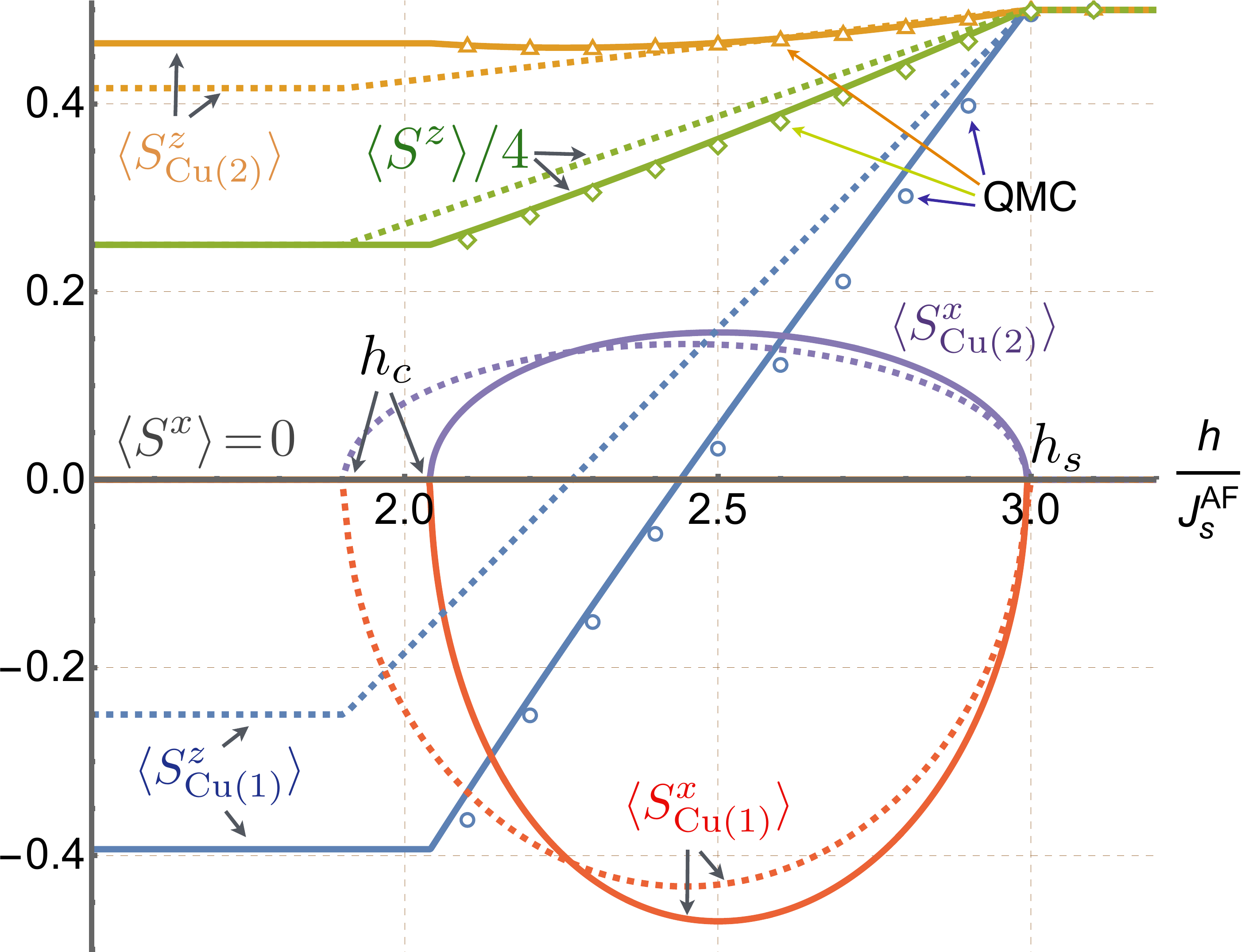}
\caption{\label{fig:TMFvsTauModel_m}
Comparison between the TMF theory (solid lines), effective $\tau$ model (dashed lines) and QMC calculations (open symbols), using the exchange parameters of Ref.~\cite{Ozerov2014}.   Results are shown for the $x$- and $z$-components of the Cu(1) and Cu(2) spins, as well as for the total $S^x$ and $S^z$.} 
\end{figure}

In terms of the underlying Cu spins, the main features of the intermediate phase, as revealed by the $\tau$ model,  are shown by dashed lines in Fig.~\ref{fig:TMFvsTauModel_m}. Between $h_\mathrm{c}$ and $h_\mathrm{s}$, the longitudinal magnetization increases linearly with field, whereas the individual transverse components ($\langle S^x_\mathrm{Cu(1)} \rangle$ and $\langle S^x_\mathrm{Cu(2)} \rangle$) show a dome-like behavior, with the total transverse moment vanishing identically, as in the magnon BECs observed in dimerized spin-$\frac12$ systems~\cite{RevModPhys.86.563,Giamarchi2008,Jaime2004,Sebastian2006,Tanaka2001,Ruegg2003,PhysRevLett.125.267207}. Unlike these systems, however, which can be described as canted antiferromagnets, here we are dealing with a canted ferrimagnet.

By construction, the $\tau$ model does not take into account the  strong admixing between $|1,3/2,1,1\rangle$ and $|1,3/2,2,1\rangle$, which plays a significant role at low fields~\cite{Janson2014,Judit2014} and is expected to survive up to $h>h_\mathrm{c}$. 
%
%mixing between $|1,3/2,1,1\rangle$ and $|1,3/2,2,1\rangle$,  mentioned above. 
%
To incorporate this mixing and obtain a more complete picture at the mean-field level, we now turn to the so-called tetrahedral mean-field (TMF) theory~\cite{Janson2014}. In this approach, one treats the strong intra-tetrahedral interactions fully quantum-mechanically, whereas the weaker inter-tetrahedral couplings are treated at a mean-field level. The TMF Hamiltonian takes the form
\be\label{eq:Htmf}
\mc{H}_{\text{TMF}} = \mc{H}_0 + \zeta_z {S}_{123}^z +\eta_z {S}_{4}^z + \zeta_x {S}_{123}^x +\eta_x {S}_{4}^x\,,
\ee
with the self-consistent mean-field parameters 
\be
\ra{1.5}
\begin{array}{l}
\bs{\zeta} = 2 \jwfm \langle {\bf S}_1\rangle + (\jwaf + \jooaf) \langle {\bf S}_4\rangle\,,
\\
\bs{\eta} = 3 (\jwaf + \jooaf)  \langle {\bf S}_1\rangle\,,
\end{array}
\ee
and we choose $\zeta^y\!=\!\eta^y\!=\!0$. 
Compared to the TMF theory developed for low-fields~\cite{Janson2014}, here the last two terms of Eq.~(\ref{eq:Htmf}) are new and become relevant in the intermediate phase. In the absence of these last two terms, one needs to diagonalize $\mc{H}_{\text{TMF}}$ in the  basis formed by the states $|1,3/2,1,1\rangle$ and $|1,3/2,2,1\rangle$. By contrast, in the intermediate phase, where $\zeta_x$ and $\eta_x$ are nonzero, one must diagonalize $\mc{H}_{\text{TMF}}$ in a larger, $8\times8$ basis, formed by the triplet $|1,3/2,1,M\rangle$ and the quintuplet $|1,3/2,2,M\rangle$.  

The TMF results are shown in Fig.~\ref{fig:TMFvsTauModel_m} by solid lines. 
Compared to the results delivered by the $\tau$ model (dashed lines), there are two observable differences (see also Table~\ref{tab:BcBs} and Fig.~\ref{fig:SzCu2} in End Matter):  
i) 
The behavior of $\langle S_{\text{Cu}(2)}^z\rangle$ right above $h_\mathrm{c}$ is linear in the $\tau$ model but non-monotonic (with a minimum) in the TMF theory.
ii) The value of $h_\mathrm{c}$ is  underestimated in the $\tau$ model. Both differences stem from the fact that the $\tau$ model does not include the admixing mentioned above, which remains appreciable from $h\!=\!0$ up to the vicinity of $h_\mathrm{c}$. 
Compared to experiment, the TMF critical fields are $B_\mathrm{c}\!=\!209$\,T and $B_\mathrm{s}\!=\!306$\,T, very close to the measured values of $180(5)$\,T and $300(15)$\,T, respectively.

For comparison, Fig.~\ref{fig:TMFvsTauModel_m} shows also results from Quantum Monte Carlo (QMC) simulations (symbols), on finite clusters of up to $6\times\!6\times\!6$ cells (3456 spins) with periodic boundary conditions, with $k_\mathrm{B}T = 0.03 \jsaf$ and 300\,000 sweeps per measurement. These simulations are based on the stochastic series expansion quantum Monte Carlo code \textsc{dirloop\_sse}~\cite{dirloop_sse_1, dirloop_sse_2}, which is part of the \textsc{alps} package~\cite{alps1.3, alps2.0} version 2.3.0.  
The agreement between the TMF theory and QMC is quite satisfactory, indicating that the quantum entanglement is well captured in the TMF approach, as in low fields~\cite{Janson2014}.

Finally, the TMF theory can be extended to nonzero temperatures, leading to the stability region of the magnon BEC  phase shown in Fig.~\ref{fig:H0spectrum-PD}\,(b). According to these results, the magnon BEC phase can survive up to $~0.2\jsaf \simeq 30$~K, half the zero-field ordering temperature $T_\mathrm{c}\simeq 60$~K.
Quantum fluctuations beyond the mean-field level are expected to modify the critical behavior near the $T\!=\!0$ critical points~\cite{RevModPhys.86.563,Giamarchi2008}, but otherwise the mean-field picture is sufficient for our purposes.

\begin{figure}[!b]
\centering
\includegraphics[width=0.9\columnwidth]{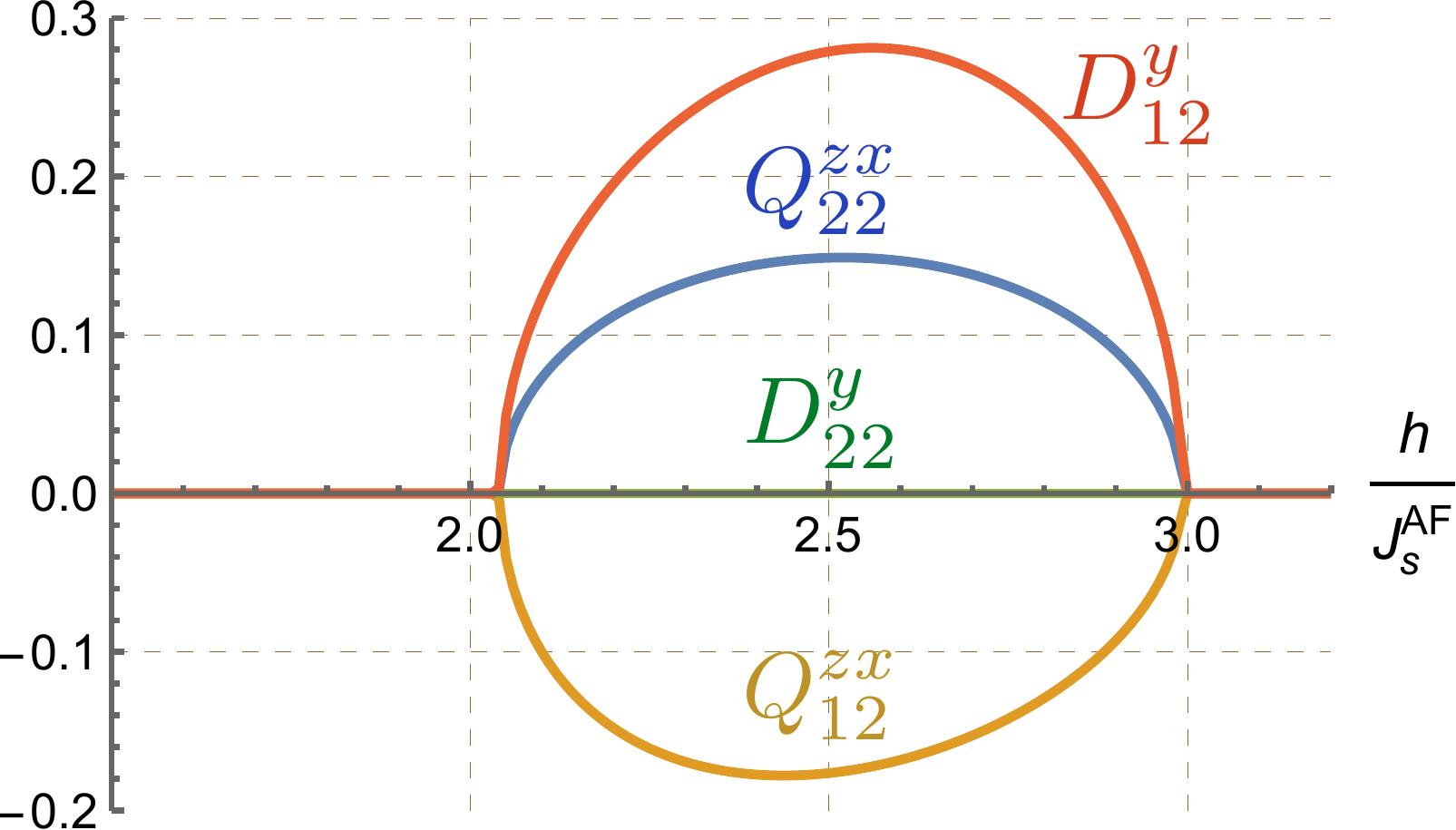}
\caption{\label{fig:TMFQD}
Zero-temperature quantum-mechanical expectation values of the operators $Q^{zx}_{ij}\!=\!\langle 
S^z_{\text{Cu(i)}}S^x_{\text{Cu(j)}}\!+\!S^x_{\text{Cu(i)}}S^z_{\text{Cu(j)}}\rangle$ and $D^{y}_{ij}\!=\!\langle  S^z_{\text{Cu(i)}}S^x_{\text{Cu(j)}}\!-\!S^x_{\text{Cu(i)}}S^z_{\text{Cu(j)}}\rangle$ ($i,j\!=\!1,2$), in the TMF ground state, for the exchange parameters of Ref.~\cite{Ozerov2014}, for fields along $z$ and transverse moments along $x$.} 
\end{figure}

{\it Origin of Faraday rotation.} 
Let us now turn to the Faraday rotation angle $\theta_\mathrm{F}'$ (Fig.~\ref{fig:fr}\,(b)). % with the input from the theoretical calculations. 
In the intermediate phase, $\theta_\mathrm{F}'$ includes the linear magnetization term $\theta_\mathrm{MOA}$ (dashed line) and the dome-like component $\theta_\mathrm{D}$. A reasonable hypothesis is that the former is driven by the longitudinal magnetization~\cite{PhysRevB.94.094409,PhysRevB.95.235155}, whereas the latter is driven by the individual transverse spin components $\langle S^x_\mathrm{Cu(1)} \rangle$ and $\langle S^x_\mathrm{Cu(2)}\rangle$. Given that the total transverse moment vanishes identically, this points to the so-called electric field-induced Faraday rotation mechanism (EFIF)~\cite{PhysRevApplied.5.031001,Krichevtsov1988}, which is allowed in systems that break both time- and space-inversion symmetry. Here it is associated with the presence of a nonzero electric polarization in the BEC phase. %The EFIF effect is allowed in systems that break both the time and space-inversion symmetry.

To check that such a polarization is indeed present, we follow the phenomenological theory of Ref.~\cite{PhysRevB.85.224413} and write the polarization ${\bf P}$ as a sum of contributions from neighboring Cu-Cu sites $(i,j)$,
\be\label{eq:Pij}
P_{ij}^\alpha = \frac{\chi_e}{2} \left(
\xi_+\big{\langle} S^\beta_iS^\gamma_j + S^\gamma_iS^\beta_j\big{\rangle}
 + \xi_-\big{\langle} ({\bf S}_i \times {\bf S}_j)^\alpha\big{\rangle}\right),~~
\ee  
where $\chi_e$ is the electric susceptibility, $(\beta,\gamma)=(y,z)$, $(z,x)$ and $(x,y)$ for $\alpha=x$, $y$ and $z$, respectively, and $\xi_+$ and $\xi_-$ are ME coupling constants, associated with the quadrupolar and the DM mechanism, respectively. 
For low fields, the latter is ineffective as spins are nearly collinear at short distances~\cite{PhysRevB.85.224413}, but this is not the case in the BEC
%intermediate canted 
phase for Cu(1)-Cu(2) pairs. 

The direction and magnitude of ${\bf P}_{ij}$ depend on the direction of ${\bf B}$ and that of the transverse order parameter.  
For example, for ${\bf B}$ along $z$ and transverse moments along $x$ (as in  Fig.~\ref{fig:TMFvsTauModel_m}), ${\bf P}_{ij}$ points along $y$ with $P_{ij}^y\!\propto\!\langle S_{\text{Cu(i)}}^z S_{\text{Cu(j)}}^x\rangle$, leading to a slightly asymmetric, dome-like contribution (see Fig.~\ref{fig:TMFQD}). 
Such contributions arise also for other field directions, as can be checked by using Eq.~(\ref{eq:Pij}).

{\it Discussion.}
The semi-quantitative agreement between experiment and theory up to hundreds of Tesla suggests that the change of the exchange parameters with magnetic fields is relatively small.
This stands in sharp contrast to geometrically frustrated pyrochlore compounds~\cite{PhysRevLett.93.197203,PhysRevLett.107.207203,PhysRevB.101.054432,doi:10.1073/pnas.2302756120}, where spin–lattice coupling plays an essential role in lifting the degeneracy.
In Cu$_2$OSeO$_3$, geometric frustration is negligible because antiferromagnetic and ferromagnetic bonds coexist, stabilizing the $\frac12$ plateau even in the absence of lattice distortions.

Likewise, the theory accounts for all basic features without the need to incorporate the DM anisotropy. While the latter is appreciable~\cite{PhysRevLett.109.107203,Janson2014} and plays a crucial role at low fields, its impact at high fields is extremely small, as can be seen by incorporating the DM interactions into the effective $\tau$ model (see SM~\cite{suppl}). Specifically, the effective couplings resulting from the DM anisotropy include: local Zeeman fields (which, for ${\bf B}\parallel{\bf e}_z$,  point along $\pm ({\bf e}_ x\pm{\bf e}_y)$, which become the `easy axes' for the transverse moments), plus effective antisymmetric (with ${\bf D}_{\text{eff}}\!\parallel\!{\bf B}$) and symmetric exchange couplings between the $\tau$'s, which can give rise to weak indentations in the magnetic order (such as further canting or long-wavelength incommensurate superstructures). The effective couplings are, however, too weak ($\lesssim 0.5$~K) to be relevant for the interpretation of the experimental data.

{\it Conclusion.} 
Using ultrahigh magnetic fields up to 500~T, we have observed that the disentangling of the Cu$_4$ building blocks of Cu$_2$OSeO$_3$ proceeds via a canted XY ferrimagnetic phase, sandwiched between the $\frac12$ plateau ($B_\mathrm{c}=180$\,T) and the fully polarized phase ($B_\mathrm{s}=300$\,T). This intermediate phase can be thought of as a BEC of a ${\bf Q}\!=\!0$ triplet-quintuplet mode.

The observation of the magnon BEC phase was made possible by two crucial factors. 
First, by the use of ultrahigh magnetic fields, large enough to overcome the $\frac12$ plateau, which is one of the largest ever observed~\cite{PhysRevLett.107.207203,PhysRevB.101.054432,doi:10.1073/pnas.2302756120,PhysRevLett.111.137204,PhysRevB.98.020404}. 
Second, the presence of linear ME coupling is also crucial. Indeed, the long-range coherence in the plane perpendicular to the field ${\bf B}$ gives rise to a slightly asymmetric dome-like contribution to the electric polarization (similarly to Ref.~\cite{Kimura2016}). In turn, this gives rise to a similar contribution to the Faraday rotation angle, via the EFIF mechanism~\cite{PhysRevApplied.5.031001,Krichevtsov1988}. So, the long-range order, which is otherwise difficult to observe due to the vanishing of ${\bf M}_\perp$, is revealed via the ME coupling. Our study broadens the spectrum of physical properties known in Cu$_2$OSeO$_3$ and highlights the rich interplay between strong electron-electron interactions, magnetoelectric coupling, and strong magnetic fields.

%\clearpage

\section*{acknowledgments}
TN thank H. Sawabe for the experimental supports and M. Matsumoto for fruitful discussions.
This work was partly supported by JSPS KAKENHI, Grant-in-Aid for Scientific Research (Nos. JP20K14403, JP21H04990, JP22H04965, JP23H04859, JP23H04861, JP24H02235, JP25H00611), 
JST CREST (grant no JPMJCR23O4), Asahi Glass Foundation, and Murata Science Foundation. 
IR acknowledges the support by the Engineering and Physical Sciences Research Council, Grant No.
EP/V038281/1. 
OJ acknowledges the support by the German Forschungsgemeinschaft (DFG, German Research Foundation) through SFB 1143 (Project ID 247310070) and thanks Ulrike Nitzsche for technical assistance.

%\section*{Data availability}
%The supporting data for this article
%are openly available from Zenodo {\cred \cite{}}

\bibliography{b1}

\clearpage 
\section*{End Matter}

\textit{Comparison between various exchange parameter sets--}
Table~\ref{tab:set} lists five sets of exchange parameters reported in the literature. 
Table~\ref{tab:BcBs} gives the critical fields obtained from the $\tau$ model, the TMF model, QMC calculations, and the experiment. It also contains the TMF prediction for the zero-field ordering temperature  $T_\mathrm{c}^{\text{TMF}}$. 

We note the following: i) The parameters of \cite{PhysRevLett.109.107203} underestimate all quantities appreciably. 
ii) The parameters of \cite{Tucker2016} overestimate $T_\mathrm{c}^{\text{TMF}}$, which is likely due to the much larger value of $\jooaf$ compared to {\it ab initio} estimates~\cite{Janson2014,PhysRevLett.109.107203}). 
iii) The parameters of \cite{Janson2014} overestimate both critical fields, since $\jsaf$ is overestimated~\cite{Ozerov2014}.
iv) The parameter sets of \cite{Ozerov2014} and \cite{Portnichenko2016} give the same critical fields since these two sets differ only in the value of $\jsfm$ (the Cu(2)-Cu(2) coupling inside the Cu$_4$ clusters),  which does not affect the critical fields (as the three Cu(2) spins form a triplet). 
v) $h_\mathrm{c}$ is consistently lower in the $\tau$ model compared to the TMF results (see main text). 

Overall, the experimental values agree better with the parameters of Refs.~\cite{Ozerov2014,Portnichenko2016}. 
In these works, the values of the strong exchange couplings have been refined compared to those in \cite{Janson2014}, based on high-field ESR data up to 64~T~\cite{Ozerov2014} and inelastic neutron scattering data~\cite{Portnichenko2016}. By contrast, the weak couplings, which affect the low-energy physics at low fields, have been kept the same as in \cite{Janson2014}, to retain the agreement with low-field experiments.

\vspace{5mm}
\textit{Physics of an isolated Cu$_4$ cluster--}
The Hamiltonian of an isolated Cu$_4$ cluster (see Eq.~(\ref{eq:H0}) of the main text) can be re-written as
\bea
\mc{H}_0&\!=\!&
\frac{1}{2} \jsaf {\bf S}^2 + \frac{1}{2}(\jsfm\!-\!\jsaf) {\bf S}_{123}^2-h S^z + c\,,
\eea
where $c=-\frac{3}{8}(\jsaf+3\jsfm)$ is an overall  constant. Thus, the eigenstates of $\mc{H}_0$ can be labeled by $|S_{12},S_{123},S,M\rangle$, with corresponding eigenenergies given by
\bea
\!\!\!E_0(\!\!\!\!\!\!\!\!\!\!\!&\!\!\!\!\!\!\!\!\!S_{12},S_{123},S,M) = \frac{1}{2}\jsaf S(S+1) \nonumber\\
&
~~+ \frac{1}{2}(\jsfm-\jsaf) S_{123}(S_{123}+1) - h M + c\,,
\eea
which are independent of $S_{12}$. Using the rules of addition of angular momenta we obtain two singlets ($|0,1/2,0,0\rangle$ and $|1,1/2,0,0\rangle$), three triplets ($|1,3/2,1,M\rangle$, $|0,1/2,1,M\rangle$  and $|1,1/2,1,M\rangle$), and one quintuplet ($|1,3/2,2,M\rangle$), with energies (measured from that of the state $|1,3/2,1,0\rangle$,   $\Delta E \equiv E-E(1,3/2,1,0)$): 
\be
\ra{1.3}
\begin{array}{lcl}
\Delta E(1,3/2,1,M) &=&-h M\,,\\ 
\Delta E(0,1/2,0,0) &=& (\jsaf-3\jsfm)/2\,,\\ 
\Delta E(1,1/2,0,0) &=& (\jsaf-3\jsfm)/2\,,\\
\Delta E(1,3/2,2,M) &=& 2\jsaf -h M\,,\\
\Delta E(0,1/2,1,M) &=& 3(\jsaf-\jsfm)/2-h M\,,\\
\Delta E(1,1/2,1,M) &=& 3(\jsaf-\jsfm)/2-h M\,.
\end{array}
\ee
The evolution of these energies with $h/\jsaf$ is shown in Fig.~\ref{fig:H0spectrum-PD}\,(a) of the main text.

\begin{table}[bth]
\caption{\label{tab:set} Sets of exchange parameters (in units of K) reported in the literature.}
\begin{ruledtabular}
\begin{tabular}{lccccc}
\!\!$
\begin{array}{l}
\text{Parameter}\\
\text{set}
\end{array}
$
& $\jsaf$ & $\jsfm$ & $\jwaf $ & $\jwfm$ & $\jooaf$ \\
\hline
Ref.~\cite{PhysRevLett.109.107203} & 75.8 & $-42.9$ & 10.4 & $-13.1$ & $11.4$
\\
Ref.~\cite{Tucker2016}& $135$ & $-157$ & $4.8$ & $-42$ & $91$ 
\\
Ref.~\cite{Janson2014} & 170 & $-128$ & 27 & $-50$ & $45$ \\
Ref.~\cite{Ozerov2014}& 145 & $-140$ & 27 & $-50$ & $45$ \\
Ref.~\cite{Portnichenko2016}& 145 & $-170$ & 27 & $-50$ & $45$ \\
\end{tabular}
\end{ruledtabular}
%\end{table}

%\begin{table}[!t]
\centering
\caption{Critical fields $B_\mathrm{c}$, $B_\mathrm{s}$, and their difference $\Delta B$, as obtained from the $\tau$ model, the TMF theory, and QMC calculations using the parameter sets of Table~\ref{tab:set} and $g=2.11$~\cite{LARRANAGA20091,Belesi_2011,Ozerov2014}. 
The last column gives the TMF prediction for the zero-field ordering temperature. 
For comparison, we also show the experimental values with uncertainty ranges.}

\begin{ruledtabular}
\begin{tabular}{llccc|c}
%\!\!$\begin{array}{l}
%\text{Parameter}\\ 
%\text{set}
%\end{array}$
Parameter& & $B_\mathrm{c}$ & $B_\mathrm{s}$ & $\Delta B$ & $T_\mathrm{c}^{\text{TMF}}$\\
set& & (T) & (T) & (T) & (K)\\
%& & $B_\mathrm{c}$ (T) & $B_\mathrm{s}$(T) & $\Delta B$(T) & $T_\mathrm{c}^{\text{TMF}}$ (K)\\
 \hline
\multirow{2}{*} {Ref.~\cite{PhysRevLett.109.107203}} & $\tau$ model&  105 & 138 & 33 & \\
 & TMF & 107  & 138  & 31 & 22--25\\
 \hline
 \multirow{2}{*}{Ref.~\cite{Tucker2016}} 
& $\tau$ model&  185 & 326 & 141 & \\
& TMF & 207 & 326 & 119 & 97\\ 
\hline
\multirow{3}{*} {Ref.~\cite{Janson2014}} & $\tau$ model& 230 & 341 & 111 &\\
 & TMF & 242 & 341 & 99 & 85 \\
 & QMC & 246 & 342 & 96 & \\
\hline
\multirow{3}{*}{\shortstack{Refs. \\ \cite{Ozerov2014,Portnichenko2016}}} 
& $\tau$ model& 195 & 306 & 111 &\\
 & TMF & 209 & 306 & 97 & 85--86\\
 & QMC & 214 & 305 & 91 &\\ 
 \hline
\multicolumn{2}{l}{Experiment} & 180(5) & 300(15) & 120(20) & 60~\cite{PhysRevB.78.094416} \\
\end{tabular}
\end{ruledtabular}
\label{tab:BcBs}
\end{table}

\begin{figure}[!tbh]
\centering
\includegraphics[width=0.8\columnwidth]{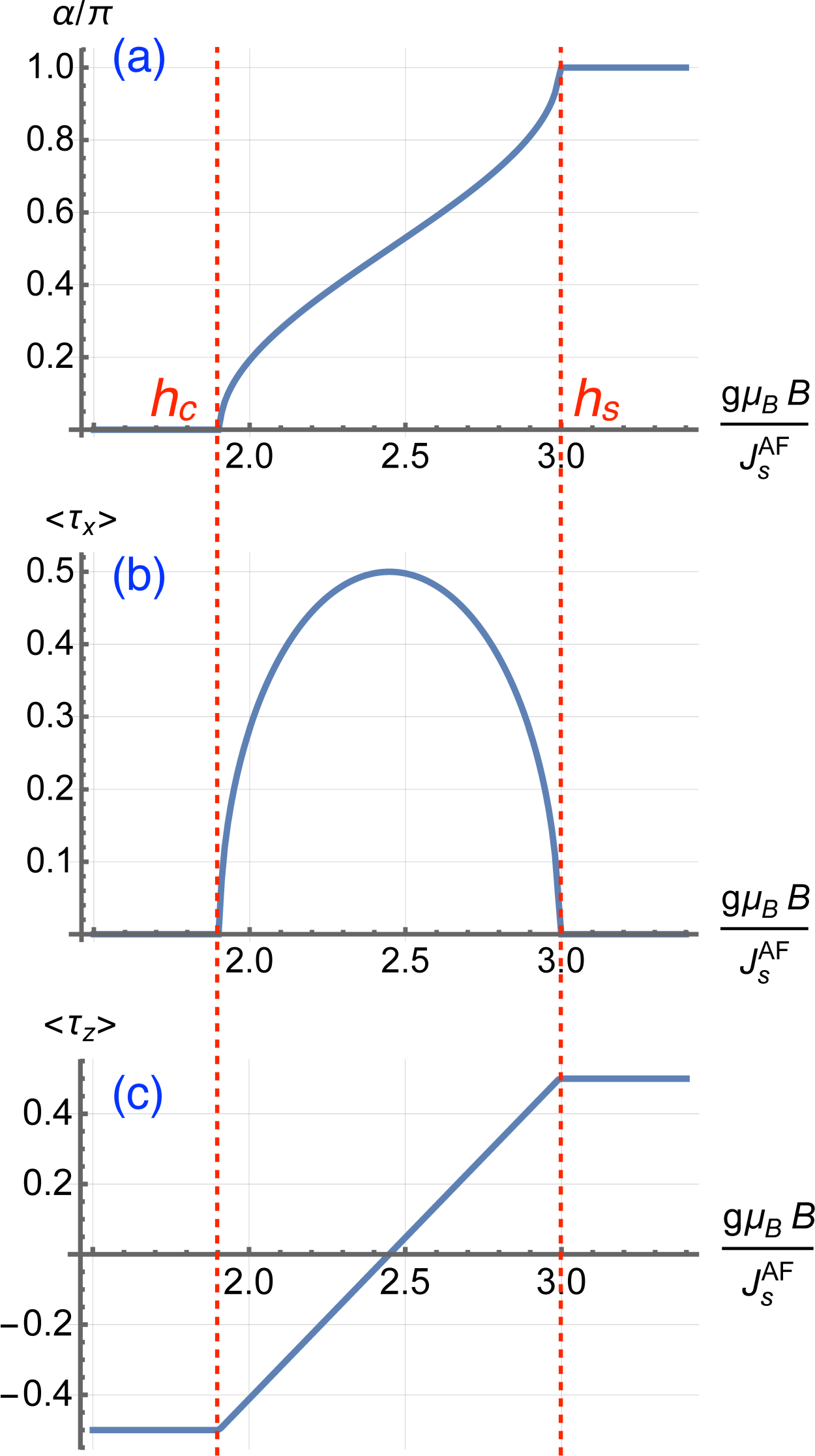}
\caption{\label{fig:TauModelsol}
%Zero-temperature 
Variational solution of the $\tau$ model for the parameters of Ref.~\cite{Ozerov2014}
: Field evolution of the variational parameter $\alpha$ (a), and the expectation values of $\tau^x$ (b) and $\tau^z$ (c).} 
\end{figure}

\begin{figure}[!tbh]
\includegraphics[width=0.8\columnwidth]{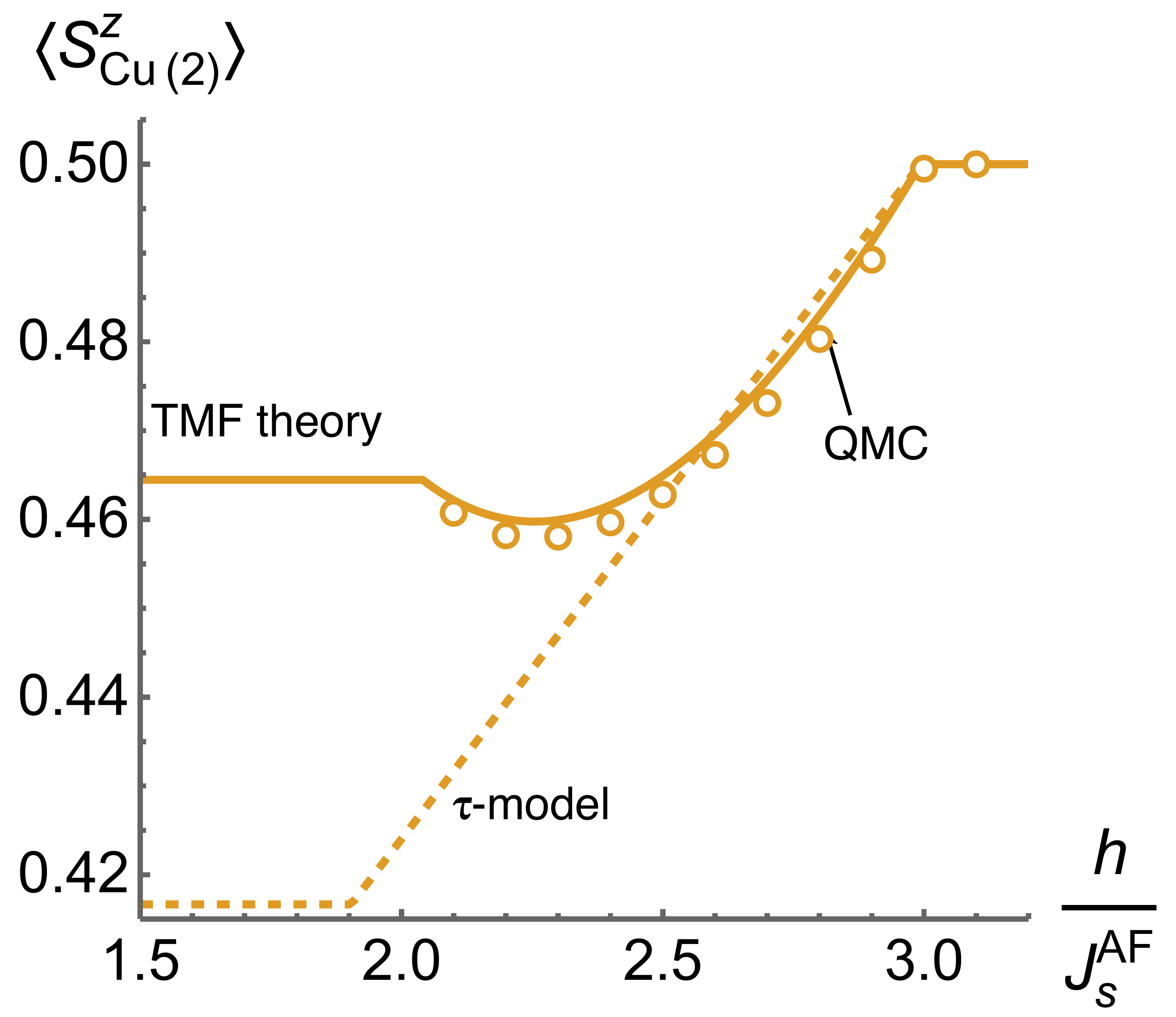}
\caption{\label{fig:SzCu2} Zoom-in on the behavior of $\langle S^z_{\text{Cu(2)}}\rangle$ shown in Fig.~\ref{fig:TMFvsTauModel_m} of the main text. TMF theory is shown with solid lines, the effective $\tau$ model with dashed lines, and QMC with symbols.} 
\end{figure}

\vspace{5mm}
\textit{Effective pseudospin-$\frac12$ model--}
The mapping of Eq.~(\ref{eq:taumapping}) of the main text leads to the equivalent operators
\be\label{eq:equivops}
\ra{1.25}
\!\!\!\begin{array}{lll}
S_{1-3}^x \mapsto \frac{1}{2\sqrt{3}} \tau^x,
\!&\!S_{4}^x \mapsto -\frac{\sqrt{3}}{2} \tau^x,
\!&\!S^x\mapsto 0,\\
S_{1-3}^y \mapsto \frac{1}{2\sqrt{3}} \tau^y,
\!&\!S_{4}^y \mapsto -\frac{\sqrt{3}}{2} \tau^y,
\!&\!S^y\mapsto 0,\\
S_{1-3}^z \mapsto \frac{11}{24}+\frac{1}{12} \tau^z,
\!&\!S_{4}^z \mapsto \frac{1}{8}+\frac{3}{4} \tau^z,
\!&\!S^z\mapsto \frac{3}{2}+\tau^z,
\end{array}
\ee
and $\mc{H}_0 \mapsto - (h-2\jsaf) \tau^z$, modulo an overall constant. 
Using these mappings one can obtain an effective Hamiltonian that describes the interactions between the $\tau$ pseudospins. Apart from an overall constant, this reads
\be
\!\mc{H}_{\text{eff}} \!=\!\!
\sum_{n=1}^2\!\sum_{\langle tt'\rangle_n}\!\!J_{n}^{xy}(\tau_t^x \tau_{t'}^x+\tau_t^y \tau_{t'}^y) + J_{n}^z \tau_t^z \tau_{t'}^z
-\widetilde{h} 
\sum_{t}\!\! \tau_t^z,
\ee
where $\langle tt'\rangle_1$ and $\langle tt'\rangle_2$ denotes first- and second-neighbor Cu$_4$ clusters, respectively, and the effective parameters 
\bea
&& 
J_{1}^{xy} = -\frac{1}{4}(\jwaf+\jooaf)\,,~~
J_{2}^{xy} = \frac{1}{12}\jwfm\,,\\
&&
J_{1}^{z} = \frac{1}{16}(\jwaf+\jooaf)\,,~~
J_{2}^{z} = \frac{1}{144}\jwfm\,,\\
&&
\widetilde{h}=
h-2\jsaf 
-\frac{11}{48}\jwfm-\frac{17}{16}(\jwaf+\jooaf)\,.
\eea
%As required, the effective model inherits the U(1)  symmetry of the original model, and is this symmetry that breaks down spontaneously in the {\cbl magnon BEC} phase.
%
Apart from the coupling $J_{1}^{z}$ (which is four times weaker than $J_{1}^{xy}$), all other couplings are ferromagnetic. So we anticipate that the $\tau$ spins order ferromagnetically in the intermediate phase.
We also note that both $S^x$ and $S^y$ map to $0$ in the effective picture. 
Hence, the total transverse magnetization vanishes identically.
%, which is different from other magnon BEC systems~\cite{Giamarchi2008,Samulon2009,Watanabe2023,PhysRevB.107.144427}.

\vspace{5mm}
\textit{Variational solution of the $\tau$ model--}
In the absence of DM anisotropy, we anticipate that the intermediate phase is uniform. We can then try a variational mean-field calculation using a state where all $\tau$ pseudospins point along the same direction, somewhere in the $xz$ plane (we pick the $x$ axis for the transverse components of the spins), as $
|\psi\rangle = \prod_t^\otimes |\alpha\rangle_t$, 
with
\be
|\alpha\rangle_t = \sin\frac{\alpha}{2}~|\tau_t^z=\frac{1}{2}\rangle + \cos\frac{\alpha}{2}~|\tau_t^z=-\frac{1}{2}\rangle\,.
\ee
%We expect the angle $\alpha$ to interpolate between $\alpha\!=\!0$ ($\langle\tau^z\rangle\!=\!-1/2$ and $\langle S^z\rangle\!=\!1$) at $h\!=\!h_\mathrm{c}$ and $\alpha\!=\!\pi$ ($\langle\tau^z\rangle\!=\!\frac12$ and $\langle S^z\rangle\!=\!2$) at $h\!=\!h_\mathrm{s}$. 

Taking into account that each $\tau$ pseudospin is coupled to six first-neighbor $\tau$'s and six second-neighbor $\tau$'s~\cite{Janson2014} we find that, apart from an overall constant, the variational energy is proportional to
$
E(\alpha) \propto c_0 + 4 c_1 \cos(\alpha) + c_2 \cos(2\alpha)$, 
where 
\be
\ra{1.25}
\!\!\!\begin{array}{l}
c_0=-27 (\jooaf + \jwaf) + 13 \jwfm\,,
\\
c_1= 
48~h_{\text{eff}}\,,~~
c_2= 45 (\jooaf + \jwaf) - 11 \jwfm \,.   
\end{array}
\ee
Minimizing with respect to $\alpha$ gives $\sin\alpha ~( c_1 + c_2 \cos\alpha) = 0$, which, in turn, gives $\sin\alpha=0$ or $\cos\alpha = -c_1/c_2$. The first solution leads to $\alpha=0$ for $h\leq h_\mathrm{c}$ and $\alpha=\pi$ for $h\geq h_\mathrm{s}$. 
The second solution is valid (and lower in energy than the first) if $|\cos\alpha|\leq1$ which happens for $h_\mathrm{c}\leq h \leq h_\mathrm{s}$. The critical fields can be found by the conditions
%\be
$\cos\alpha = \pm 1$, 
%\Rightarrow c_1 = \mp c_2\,,
%\ee
which lead to Eqs.~(\ref{eq:hchs}). 

The evolution of $\alpha$, $\langle\tau^z\rangle=-\frac{1}{2}\cos{\alpha}$ and $\langle\tau^x\rangle=\frac{1}{2}\sin\alpha$ with $h$ are shown in Fig.~\ref{fig:TauModelsol}, and are consistent with the picture of a pseudospin rotating from $-z$ to $+z$ in the $xz$ plane. 
Indeed, $\alpha$ evolves from $0$ at $h\!=\!h_\mathrm{c}$ to $\pi$ at $h\!=\!h_\mathrm{s}$,  %the transverse polarization 
$\langle\tau^x\rangle$ shows a dome-shape behavior, and %the longitudinal polarization 
$\langle\tau^z\rangle$ shows a linear increase from $-\frac12$ at $h\!=\!h_\mathrm{c}$ to $+\frac12$ at $h\!=\!h_\mathrm{s}$.

\vspace{5mm}
\textit{Behavior of $\langle S^z_{\text{Cu(2)}}\rangle$ with field--}
As mentioned in the main text, the behavior of $\langle S^z_{\text{Cu(2)}}\rangle$ right above the critical field $h_\mathrm{c}$ is qualitatively different in the $\tau$ and the TMF model, due to the fact that the former model does not take into account the mixing between $|1,3/2,1,1\rangle$ and $|1,3/2,2,1\rangle$. 
To highlight this difference, Fig.~\ref{fig:SzCu2} zooms in on the corresponding curve shown in Fig.~\ref{fig:TMFvsTauModel_m}. $\langle S^z_{\text{Cu(2)}}\rangle$ shows a linear increase in the $\tau$ model, whereas it is non-monotonic in the TMF model, with a minimum at some field above $h_\mathrm{c}$.

%\end{document}

%--------------------%--------------------%--------------------%--------------------%--------------------
\clearpage

\setcounter{page}{1}
\title{Supplemental Material of 
`Quintuplet condensation in the skyrmionic insulator Cu$_2$OSeO$_3$ at ultrahigh magnetic fields'}
\maketitlesup

\section{Experimental details}
We measured single crystals of Cu$_2$OSeO$_3$ grown by the chemical vapor transport method~\cite{PhysRevB.82.144107}.
We determined the chirality of each crystal (L or D) by measuring the sign of natural optical activity at a wavelength of 1310 nm.
We polished the (110) surfaces of the D-type crystal down to the thickness of $d \sim 0.6$~mm with the transmission area of $\sim$4~mm$^2$ and fixed it inside the He-flow cryostat made of glass epoxy~\cite{PhysRevResearch.2.033257}.
We wound the pickup coil around the crystal to measure the pulsed magnetic field.
We measured the sample temperature immediately before the experiment using an E-type thermocouple.
Experiments were performed in the Faraday geometry with the field along [110] direction.

In this study, we used two types of pulsed magnets, the single-turn coil (STC) and the electromagnetic flux compression (EMFC) in Kashiwa, UTokyo~\cite{Miura2003,10.1063/1.5044557}.
We performed one EMFC experiment up to 500 T at $\sim5$~K, and several STC experiments up to 180~T to see the temperature dependence.
The field waveforms of the STC and EMFC are shown in Fig.~S\ref{fig:field}
The field duration above 100~T was of the order of $\mu$s in both techniques.
Since the EMFC allowed the experiment only for the field-increasing process, the effect of hysteresis could be verified only by using the STCs.
Within our experimental resolution, the hysteresis was not observed up to 180~T.

\renewcommand{\figurename}{Fig. S}
\setcounter{figure}{0}
\begin{figure}[b]
\centering
\includegraphics[width=8.4cm]{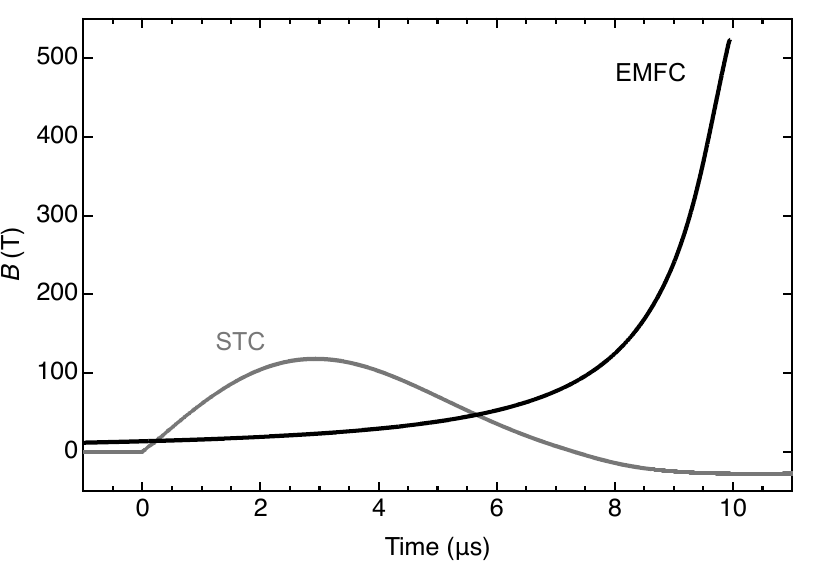}
\caption{\label{fig:field}
Magnetic-field waveforms by the single-turn coil (STC) and electromagnetic flux compression (EMFC).
} 
\end{figure}

\begin{figure}[tb]
\centering
\includegraphics[width=8.4cm]{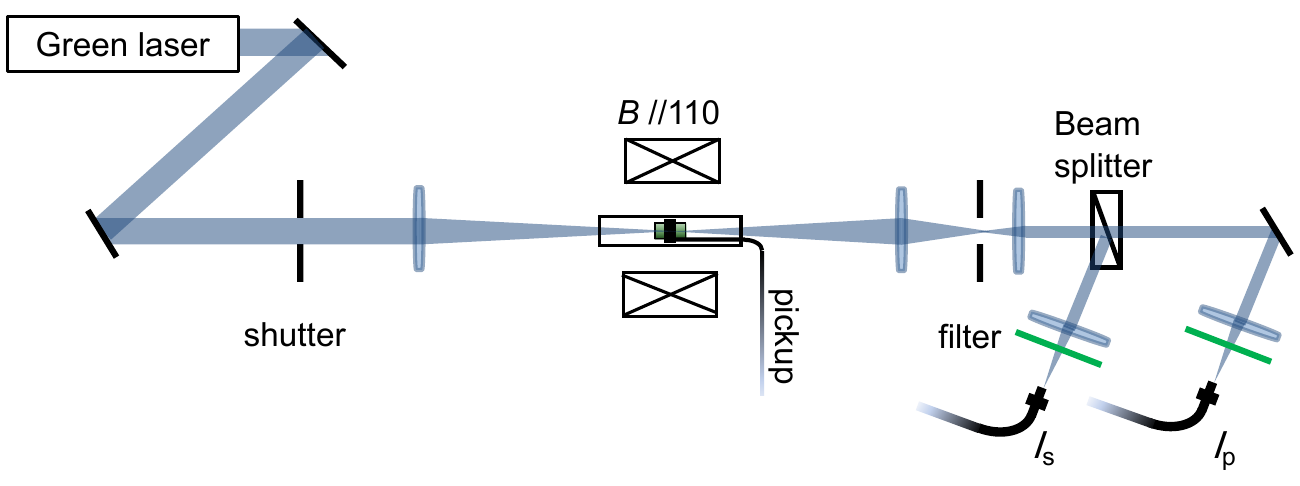}
\caption{\label{fig:exp}
Magnetic-field waveforms by the single-turn coil (blue, inset) and the dual-pulse magnet (black).
} 
\end{figure}

We performed the Faraday rotation experiments with an applied field along the [110] axis of the single crystal.
The schematic experimental setup is shown in Fig.~S\ref{fig:exp}. 
We used a green laser (Coherent Inc. Sapphire 532 nm) as an incident light source with linear polarization.
We analyzed the splitted light intensities of the s- and p-polarized components ($I_\mathrm{s}$ and $I_\mathrm{p}$).
The split lights were filtered to eliminate stray light due to the explosive field generation and transferred to silicon diode detectors via optical fibers.
$I_\mathrm{s}$, $I_\mathrm{p}$, and pickup voltage proportional to $dB/dt$ were recorded by an oscilloscope.
We calculated the Faraday rotation angle $\theta_\mathrm{F}$ using the formula 
$\theta_\mathrm{F} = \arccos{(I_\mathrm{p}-I_\mathrm{s})/(I_\mathrm{p}+I_\mathrm{s})}$.
The polarization of the transmitted light rotated even without magnetic fields because of the natural optical activity.
Here the rotation direction depends on the chirality, L- or D-type crystals.
Since the rotation angle at zero field agreed with the literature~\cite{PhysRevB.94.094409}, our single crystals were homochiral.

\section{EMFC results}

\begin{figure*}[tb]
\centering
\includegraphics[width=17.4cm]{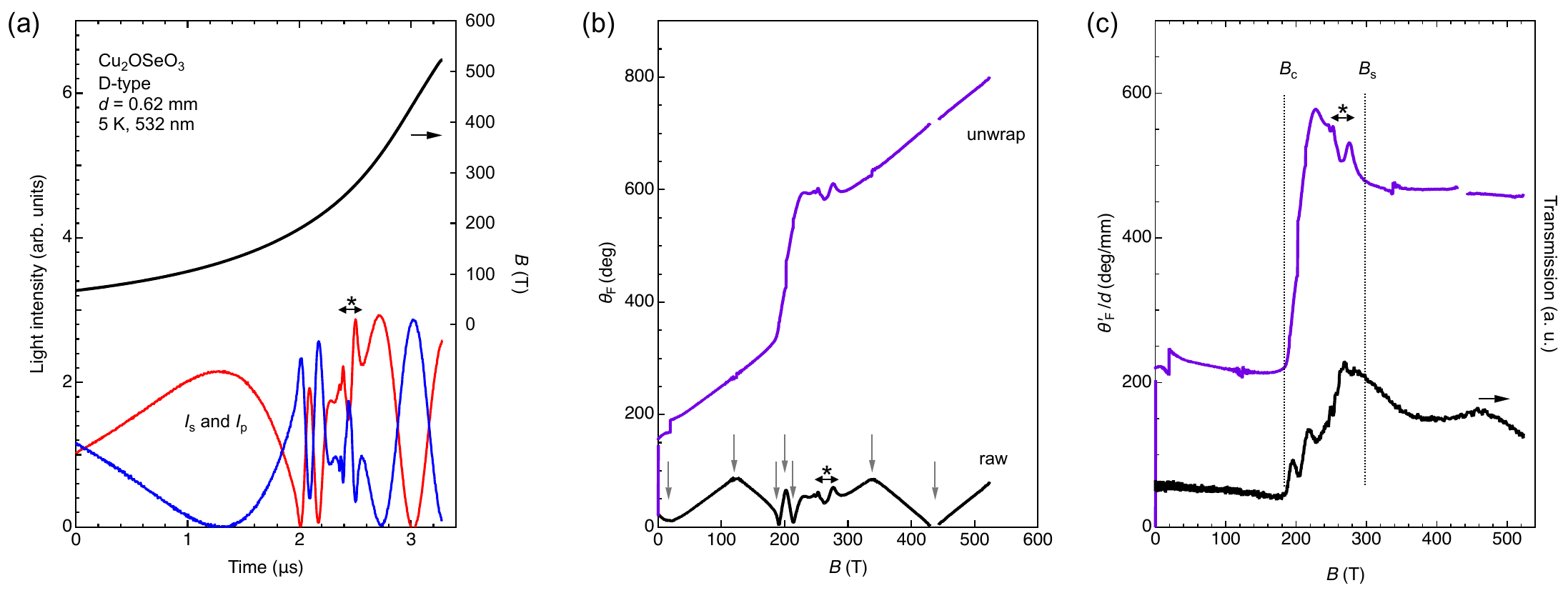}
\caption{\label{fig:emfc}
Faraday rotation results on the D-type Cu$_2$OSeO$_3$ at 5 K up to 500 T.
(a) Transmitted light intensities ($I_\mathrm{s}$ and $I_\mathrm{s}$) as a function of time. The magnetic field waveform is shown on the right axis.
(b) Faraday rotation angle $\theta_\mathrm{F}$ as a function of magnetic field. The black curve shows the calculated rotation angle (0--$\pi/2$). The purple curve shows the unwrapped result at the fields indicated by the gray arrows.
(c) Normalized Faraday rotation angle ($\theta '_\mathrm{F}/d$, see main text) and transmitted light intensity as a function of the magnetic field.
The asterisks indicate signal disturbance probably due to the mechanical vibration.
} 
\end{figure*}

Figure S\ref{fig:emfc} summarizes the raw and analyzed data of the EMFC experiment.
Figure S\ref{fig:emfc}\,(a) shows the transmitted s- and p-polarized light intensities ($I_\mathrm{s}$ and $I_\mathrm{p}$) as a function of time.
The magnetic field waveform is shown on the right axis.
The oscillating amplitude of $I_\mathrm{s}$ and $I_\mathrm{p}$ reflects the rotation of the linearly polarized light due to the Faraday effect.
The rapid oscillation at around 2.1~$\mu$s indicates that the magnetization starts to increase from 180~T.
Figure S\ref{fig:emfc}\,(b) shows the calculated Faraday rotation angle as a function of the magnetic field.
We note that the rapid increase of $\theta_\mathrm{F}$ below 0.2~T is estimated from the low-field experiment without explosion.
By unwrapping the raw data (black curve, from 0 to $\pi/2$), we can obtain the Faraday rotation angle $\theta_\mathrm{F}$ (purple curve).
The gray arrows indicate the angle (0 or $\pi/2$ radian) where the curve is unwrapped.
From 250~T to 280~T, $\theta_\mathrm{F}$ shows a hump indicated by the asterisk.
Most probably, this hump is caused by a mechanical vibration of the sample triggered by the rapid increase of magnetization around 200~T.
The magnetic phase transition can cause magnetostriction in the $\mu$s timescale, producing ultrasonic acoustic waves~\cite{PhysRevLett.125.177202,doi:10.1073/pnas.2110555118}.

Figure S\ref{fig:emfc}\,(c) shows $\theta '_\mathrm{F}/d$ and transmitted light intensity $I = \sqrt{I_\mathrm{s}^2+I_\mathrm{p}^2}$ as a function of the magnetic field. For the definition of $\theta '_\mathrm{F}$, see the main text. The transmitted light intensity also shows kinks at the magnetic transitions at $B_\mathrm{c}$ and $B_\mathrm{s}$. The change of transmission indicates that the optical absorption spectra change upon the magnetic phase transition. The wavelength of 532~nm (2.33~eV) is located at the edge of the charge transfer absorption~\cite{PhysRevB.94.094409}. Therefore, the change of $I$ indicates that the charge-transfer gap shifts to a higher energy side probably due to the structural change. However, as discussed in the main text, the calculated magnetization curve agrees well with the experimental results ($B_\mathrm{c}$ and $B_\mathrm{s}$), indicating that the structural change is not too drastic to change the exchange parameters more than 10 \%.

\section{STC results and temperature dependence}

\begin{figure}[tb]
\centering
\includegraphics[width=8.2cm]{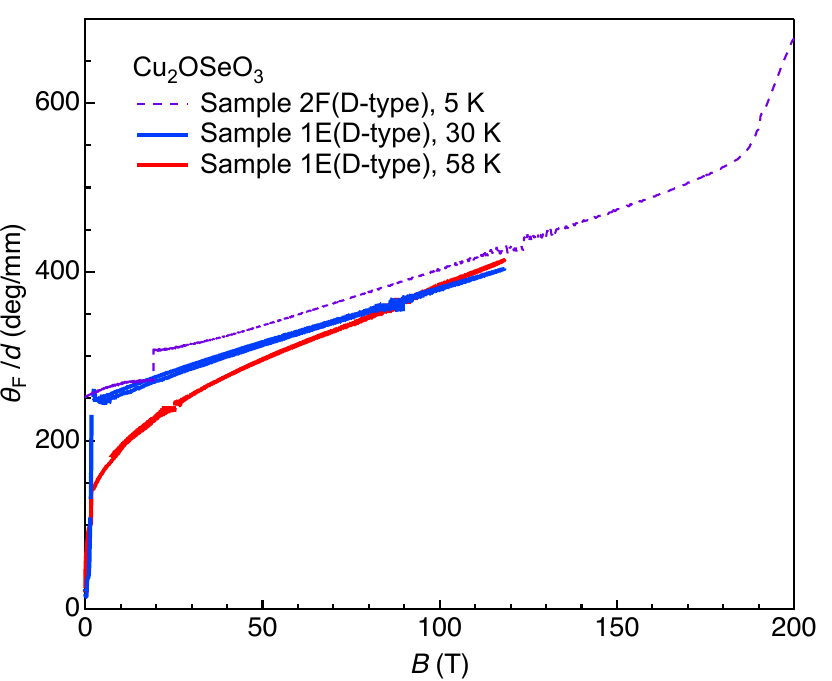}
\caption{\label{fig:stc}
Faraday rotation angle $\theta_\mathrm{F}$ as a function of magnetic field at selected temperatures.
} 
\end{figure}

Figure S\ref{fig:stc} shows the results of the STC up to 120 T, compared with the EMFC result (dotted curve). All results show the drastic increase of $\theta_\mathrm{F}$ at around 0.2~T, indicating the ferrimagnetic transition. The result at 58~K gradually approaches to the results at 30~K and 5~K since it is too close to $T_\mathrm{c}\sim58$~K. In the ferrimagnetic phase ($\frac12$ magnetization plateau), both the results at 30~K and 5~K show the same slope as a function of a magnetic field. As discussed in the main text, this linear slope is irrelevant to the magnetization since the magnetization should be constant in the $\frac12$ plateau phase. Because of the large spin gap of the order of 200 K, no temperature dependence between 30 K and 5 K is observed. Therefore, we can attribute this linear slope to the background of the Faraday rotation $\theta_\mathrm{BG}$ which is probably related to the magnetic field dependence of the optical absorption edge.

\section{Effect of Dzyaloshinskii–Moriya interactions}
To investigate the impact of the Dzyaloshinskii–Moriya (DM) interactions on the magnon BEC phase we need the list of these interactions inside the conventional unit cell and project these into the low-energy manifold of the $\tau$ model. To that end, we shall also need to differentiate between intra- and inter-tetrahedral DM couplings, as these must be treated differently in perturbation theory. 

\subsection{Structural details}
Each conventional unit cell of the system contains four Cu(1) sites (Wyckoff positions $4a$) and twelve Cu(2) sites (Wyckoff positions $12b$). 
Following Ref.~\cite{Janson2014}, we use the enantiomer defined by the following crystallographic positions  (in units of the lattice constant $8.91113~\AA$):
\be
\ra{1.25}
\begin{array}{l}
\bs{\rho}_1 = (y, y, y)\,,\\
\bs{\rho}_2 = (3/2 - y, 1 - y, y - 1/2)\,,\\
\bs{\rho}_3 = (1 - y, y - 1/2, 3/2 - y)\,,\\
\bs{\rho}_4 = (y - 1/2, 3/2 - y, 1 - y)\,,
\end{array}
\ee
for the Cu(1) sites, and
\be
\ra{1.25}
\begin{array}{l}
\bs{\rho}_5 = (a, b, c)\,,\\
\bs{\rho}_6 = (b, c, a)\,,\\
\bs{\rho}_7 = (c, a, b)\,,\\
\bs{\rho}_8 = (1 - a, b + 1/2, 3/2 - c)\,,\\
\bs{\rho}_9 = (b + 1/2, 3/2 - c, 1 - a)\,,\\
\bs{\rho}_{10} = (3/2 - c, 1 - a, b + 1/2)\,,\\
\bs{\rho}_{11} = (a + 1/2, 1/2 - b, 1 - c)\,,\\
\bs{\rho}_{12} = (1/2 - b, 1 - c, a + 1/2)\,,\\
\bs{\rho}_{13} = (1 - c, a + 1/2, 1/2 - b)\,,\\
\bs{\rho}_{14} = (1/2 - a, 1 - b, c - 1/2)\,,\\
\bs{\rho}_{15} = (1 - b, c - 1/2, 1/2 - a)\,,\\
\bs{\rho}_{16} = (c - 1/2, 1/2 - a, 1 - b)\,,
\end{array}
\ee
for the Cu(2) sites. Here $y=0.88557$, $a=0.13479$, $b=0.12096$ and $c=0.87267$.

For later purposes, we identify four strong tetrahedra ($A$-$D$) and four weak tetrahedra ($A'$-$D'$) in the conventional unit cell. The former are formed by the Cu sites 
\be
\ra{1.25}
\begin{array}{l}
A:~ \{ \bs{\rho}_4, \bs{\rho}_9 \!-\! (0, 0, 1), \bs{\rho}_{11}, \bs{\rho}_{16} \!-\! (0, 0, 1) \} \,,\\
B:~ \{ \bs{\rho}_1, \bs{\rho}_5 \!+\! (1, 1, 0), \bs{\rho}_6 \!+\! (1, 0, 1), \bs{\rho}_7 \!+\! (0, 1, 1) \}\,,\\
C:~ \{\bs{\rho}_2, \bs{\rho}_{10} \!-\! (0, 1, 0), \bs{\rho}_{12}, \bs{\rho}_{14} \!-\! (0, 1, 0) \}\,,\\
D:~ \{\bs{\rho}_3, \bs{\rho}_8 \!-\! (1, 0, 0), \bs{\rho}_{13}, \bs{\rho}_{15} \!-\! (1, 0, 0) \} \,, 
\end{array}
\ee
and the weak tetrahedra are formed by the Cu sites
\be
\ra{1.25}
\begin{array}{l}
A':~\{ \bs{\rho}_1, \bs{\rho}_8, \bs{\rho}_9, \bs{\rho}_{10} \}\,,\\
B':~\{ \bs{\rho}_2, \bs{\rho}_7, \bs{\rho}_{11}, \bs{\rho}_{15} \}\,, \\
C':~\{ \bs{\rho}_3, \bs{\rho}_5, \bs{\rho}_{12}, \bs{\rho}_{16} \}\,,\\
D':~\{ \bs{\rho}_4, \bs{\rho}_6, \bs{\rho}_{13}, \bs{\rho}_{14} \}\,.
\end{array}
\ee

\subsection{List of DM vectors in a conventional unit cell}
We shall use the {\it ab initio} DM vectors reported in Ref.~\cite{Janson2014} and general symmetry arguments to connect the DM components on symmetry related bonds,  we can generate the list of all DM vectors in a conventional unit cell. The ones inside the four strong tetrahedra ($A$-$D$) are provided in  Table~\ref{tab:DMstrong}, and the ones inside the weak tetrahedra ($A'$-$D'$) are provided in Table~\ref{tab:DMweak}.

\subsection{Effect of intra-tetrahedra DM vectors}
Projecting the DM interactions inside the strong tetrahedra onto the $2\times2$ manifold of the $\tau$ model gives rise to the following insights. 

i) The interactions related to the DM vector ${\bf D}_s^{\text{FM}}$ (and the corresponding vectors on symmetry-related bonds) map to zero identically. Thus, to leading order, ${\bf D}_s^{\text{FM}}$ does not play any role. 

ii) The interactions related to the DM vector ${\bf D}_s^{\text{AF}}$ (and its symmetry-retated ones) map to:
\be
\ra{1.25}
\begin{array}{l} 
A:~V_{{\bf D}_{s}^{\text{AF}}} \mapsto - d_s ~\bs{\tau}_A \cdot {\bf e}_+ \,, \\
B:~V_{{\bf D}_{s}^{\text{AF}}} \mapsto + d_s ~\bs{\tau}_B \cdot {\bf e}_-\,,\\
C:~V_{{\bf D}_{s}^{\text{AF}}} \mapsto - d_s ~\bs{\tau}_C\cdot {\bf e}_-\,, \\
D:~V_{{\bf D}_{s}^{\text{AF}}}\mapsto + d_s~ \bs{\tau}_D \cdot {\bf e}_+\,, 
\end{array}
\ee
where $d_s=\sqrt{\frac{2}{3}}(d_x-d_y-d_z)$
and ${\bf e}_\pm = \frac{{\bf e}_x\pm{\bf e}_y}{\sqrt{2}}$. So, to leading order, the DM interactions related to ${\bf D}_{s}^{\text{AF}}$ map to local Zeeman terms along the directions $\pm ({\bf x}\pm{\bf y})$ in the $xy$ plane. These extra Zeeman terms break the U(1) symmetry explicitly. Given the fact that the dominant effective XXZ interactions between the $\tau$ variables are ferromagnetic, the extra Zeeman terms likely act to select one of the diagonal transverse directions. Although this hypothesis needs to be checked numerically, we note that the strength of the extra Zeeman terms is extremely small, $d_s\simeq0.65$\,K (using the values calculated by DFT~\cite{Janson2014}, see also Table~\ref{tab:DMstrong}).

\subsection{Effect of inter-tetrahedra DM vectors}
To project the DM interactions inside the weak tetrahedra (equivalently, the interactions between strong tetrahedra) onto the manifold of the $\tau$ model one can use the mappings given in Eq.~(\ref{eq:equivops}) of the main text (Appendix~\ref{app:tau}). For example, the DM interaction between the Cu(1) site at position $\bs{\rho}_1$ (associated, say, with $\tau_t$) and the Cu(2) site at position $\bs{\rho}_8$ (associated, say, with $\tau_{t'}$), maps to 
%\begin{small}
\be
{\bf D}_{\bs{\rho}_1,\bs{\rho}_8}\cdot
{\bf S}_{\bs{\rho}_1} \!\times\!{\bf S}_{\bs{\rho}_8}
\mapsto 
{\bf D}_{\bs{\rho}_1,\bs{\rho}_8} \cdot 
\left(
\begin{array}{c}
-\frac{\sqrt{3}}{2}\tau_t^x\\
-\frac{\sqrt{3}}{2}\tau_t^y\\
\frac{1}{8}+\frac{3}{4}\tau_t^z
\end{array}
\right)
\!\times\!
\left(
\begin{array}{c}
\frac{1}{2\sqrt{3}}\tau_{t'}^x\\
\frac{1}{2\sqrt{3}}\tau_{t'}^y\\
\frac{11}{24}+\frac{1}{12}\tau_{t'}^z
\end{array}
\right),
\ee
\normalsize
where ${\bf D}_{\bs{\rho}_1,\bs{\rho}_8}=(w_x,w_y,w_z)$, see Table~\ref{tab:DMstrong}. When expanded, the above expression gives a combination of local Zeeman terms, antisymmetric DM terms as well as symmetric exchange terms. The strength of these terms does not exceed 0.4~K. 
Similarly, the DM interaction between the Cu(2) site at positions $\bs{\rho}_5$ and the Cu(2) site at position $\bs{\rho}_{12}$, maps to 
%\begin{small}
\be
{\bf D}_{\bs{\rho}_5,\bs{\rho}_{12}}\cdot 
\left(
\begin{array}{c}
\frac{1}{2\sqrt{3}}\tau_t^x\\
\frac{1}{2\sqrt{3}}\tau_t^y\\
\frac{11}{24}+\frac{1}{12}\tau_t^z
\end{array}
\right)
\!\times\!
\left(
\begin{array}{c}
\frac{1}{2\sqrt{3}}\tau_{t'}^x\\
\frac{1}{2\sqrt{3}}\tau_{t'}^y\\
\frac{11}{24}+\frac{1}{12}\tau_{t'}^z
\end{array}
\right),
\ee
\normalsize
where ${\bf D}_{\bs{\rho}_5,\bs{\rho}_{12}}=(w_x',w_y',w_z')$, see Table~\ref{tab:DMweak}. As above, this gives a combination of local Zeeman terms, antisymmetric DM terms as well as symmetric exchange terms, whose strength does not exceed 0.13~K. 

Altogether then, the effective terms in the $\tau$ model generated by the DM interactions either vanish identically or are extremely weak to play any observable role.

\begin{table*}[!t]
\centering
\caption{Components of the DM vectors ${\bf D}_{\text{s},ij}^{\text{AF}}$ and ${\bf D}_{\text{s},ij}^{\text{FM}}$ inside the four strong tetrahedra of a conventional unit cell. The ones highlighted in blue are the DM vectors calculated by DFT (in K)~\cite{Janson2014}\label{tab:DMstrong}}
\ra{1.3}
\begin{ruledtabular}
\begin{tabular}{c	||	lll	||	lll}
$\ra{0.85}\begin{array}{c}  \text{strong}\\ \text{tetrahedron} \end{array}$ &${\bf r}_i$ & ${\bf r}_j$ & ${\bf D}_{\text{s},ij}^{\text{AF}}$ & ${\bf r}_i$ & ${\bf r}_j$ & ${\bf D}_{\text{s},ij}^{\text{FM}}$ \\
\hline
&${\cbl\bs{\rho}_{4}}$&${\cbl\bs{\rho}_{11}}$&${\cbl\begin{array}{l}(d_x,d_y,d_z)=\\(-7.4,-9.5,1.3)\end{array}}$	&$\bs{\rho}_9-(0,0,1)$&$\bs{\rho}_{16}-(0,0,1)$&$(d_y',d_z',d_x')$\\
A&$\bs{\rho}_{4}$&$\bs{\rho}_{9}-(0,0,1)$&$(-d_y,d_z,-d_x)$		&$\bs{\rho}_{16}-(0,0,1)$&$\bs{\rho}_{11}$&$(-d_z',d_x',-d_y')$\\
&$\bs{\rho}_{4}$&$\bs{\rho}_{16}-(0,0,1)$&$(-d_z,-d_x,d_y)$		&$\bs{\rho}_{11}$&$\bs{\rho}_9-(0,0,1)$&$(-d_x',-d_y',d_z')$\\
\hline
&$\bs{\rho}_{1}$&$\bs{\rho}_{5}+(1,1,0)$&$(d_x,-d_y,-d_z)$		&$\bs{\rho}_5+(1,1,0)$&$\bs{\rho}_6+(1,0,1)$&$(-d_x',d_y',-d_z')$\\
B&$\bs{\rho}_{1}$&$\bs{\rho}_{6}+(1,0,1)$&$(-d_y,-d_z,d_x)$		&$\bs{\rho}_6+(1,0,1)$&$\bs{\rho}_7+(0,1,1)$&$(d_y',-d_z',-d_x')$\\
&$\bs{\rho}_{1}$&$\bs{\rho}_{7}+(0,1,1)$&$(-d_z,d_x,-d_y)$		&$\bs{\rho}_7+(0,1,1)$&$\bs{\rho}_5+(1,1,0)$&$(-d_z',-d_x',d_y')$\\
\hline
&$\bs{\rho}_{2}$&$\bs{\rho}_{12}$&$(d_y,d_z,d_x)$				&$\bs{\rho}_{10}-(0,1,0)$&$\bs{\rho}_{14}-(0,1,0)$&$(d_z',d_x',d_y')$\\
C&$\bs{\rho}_{2}$&$\bs{\rho}_{10}-(0,1,0)$&$(d_z,-d_x,-d_y)$	&$\bs{\rho}_{14}-(0,1,0)$&$\bs{\rho}_{12}$&$(d_x',-d_y',-d_z')$\\
&$\bs{\rho}_{2}$&$\bs{\rho}_{14}-(0,1,0)$&$(-d_x,d_y,-d_z)$		&$\bs{\rho}_{12}$&$\bs{\rho}_{10}-(0,1,0)$&$(-d_y',d_z',-d_x')$\\
\hline
&$\bs{\rho}_{3}$&$\bs{\rho}_{13}$&$(d_z,d_x,d_y)$				&${\cbl\bs{\rho}_8-(1,0,0)}$&${\cbl\bs{\rho}_{15}-(1,0,0)}$& ${\cbl\begin{array}{l}(d_x',d_y',d_z')=\\(2.1,4.8,-2.7)\end{array}}$ \\
D&$\bs{\rho}_{3}$&$\bs{\rho}_{8}-(1,0,0)$&$(-d_x,-d_y,d_z)$		&$\bs{\rho}_{15}-(1,0,0)$&$\bs{\rho}_{13}$&$(-d_y',-d_z',d_x')$\\
&$\bs{\rho}_{3}$&$\bs{\rho}_{15}-(1,0,0)$&$(d_y,-d_z,-d_x)$		&$\bs{\rho}_{13}$&$\bs{\rho}_8-(1,0,0)$&$(d_z',-d_x',-d_y')$
\end{tabular}
\end{ruledtabular}
\end{table*}

\begin{table*}[!h]
\centering
\caption{Components of the DM vectors ${\bf D}_{\text{w},ij}^{\text{AF}}$ and ${\bf D}_{\text{w},ij}^{\text{FM}}$ inside the four weak tetrahedra of a conventional unit cell, as well as the vectors 
${\bf D}_{\text{O}\cdots\text{O},ij}^{\text{AF}}$. The ones highlighted in blue are the DM vectors calculated by DFT  (in K)~\cite{Janson2014}.\label{tab:DMweak}}
\ra{1.3}
\begin{ruledtabular}
\begin{tabular}{c	||	lll		||		lll		||		lll}
$\ra{0.85}\begin{array}{c}  \text{weak}\\ \text{tetrahedron} \end{array}$ & ${\bf r}_i$ & ${\bf r}_j$ & ${\bf D}_{\text{w},ij}^{\text{AF}}$ 			& ${\bf r}_i$ & ${\bf r}_j$ & ${\bf D}_{\text{w},ij}^{\text{FM}}$ 		& ${\bf r}_i$ & ${\bf r}_j$ & ${\bf D}_{\text{O}\cdots\text{O},ij}^{\text{AF}}$ \\
\hline
&${\cbl\bs{\rho}_{1}}$ & ${\cbl\bs{\rho}_{8}}$ & ${\cbl\begin{array}{l}(w_x,w_y,w_z)=\\(-5.7,-9.3,8.8)\end{array}}$	& $\bs{\rho}_{8}$ &$\bs{\rho}_{9}$ & $(-w_x',w_y',-w_z')$			& $\bs{\rho}_{1}$ &$\bs{\rho}_{14}+(0,0,1)$ & $(-p_z,p_x,-p_y)$\\
A'&$\bs{\rho}_{1}$ & $\bs{\rho}_{9}$ & $(w_y,w_z,w_x)$				& $\bs{\rho}_{9}$ &$\bs{\rho}_{10}$ & $(w_y',-w_z',-w_x')$		& $\bs{\rho}_{1}$ &$\bs{\rho}_{15}+(0,1,0)$ & $(p_x,-p_y,-p_z)$\\
&$\bs{\rho}_{1}$ & $\bs{\rho}_{10}$ & $(w_z,w_x,w_y)$				& $\bs{\rho}_{10}$ &$\bs{\rho}_{8}$ & $(-w_z',-w_x',w_y')$		& $\bs{\rho}_{1}$ &$\bs{\rho}_{16}+(1,0,0)$ & $(-p_y,-p_z,p_x)$\\
\hline
&$\bs{\rho}_{2}$ & $\bs{\rho}_{7}$ & $(-w_z,-w_x,w_y)$				& $\bs{\rho}_{7}$ &$\bs{\rho}_{11}$ & $(w_z',w_x',w_y')$			& $\bs{\rho}_{2}$ &$\bs{\rho}_{5}+(1,0,0)$ & $(p_z,-p_x,-p_y)$\\
B'&$\bs{\rho}_{2}$ & $\bs{\rho}_{11}$ & $(-w_x,-w_y,w_z)$			& $\bs{\rho}_{11}$ &$\bs{\rho}_{15}$ & $(w_x',-w_y',-w_z')$		& $\bs{\rho}_{2}$ &$\bs{\rho}_{9}-(0,1,1)$ & $(-p_x,p_y,-p_z)$\\
&$\bs{\rho}_{2}$ & $\bs{\rho}_{15}$ & $(-w_y,-w_z,w_x)$				& $\bs{\rho}_{15}$ &$\bs{\rho}_{7}$ & $(-w_y',w_z',-w_x')$		& $\bs{\rho}_{2}$ &$\bs{\rho}_{13}$ & $(p_y,p_z,p_x)$\\
\hline
&$\bs{\rho}_{3}$ & $\bs{\rho}_{5}$ & $(-w_x,w_y,-w_z)$		& ${\cbl\bs{\rho}_{5}}$ &${\cbl\bs{\rho}_{12}}$ & ${\cbl\begin{array}{l}(w_x',w_y',w_z')=\\-(3.2,0.3,2.8)\end{array}}$		& $\bs{\rho}_{3}$ &$\bs{\rho}_{6}+(0,0,1)$ & $(-p_x,-p_y,p_z)$\\
C'&$\bs{\rho}_{3}$ & $\bs{\rho}_{12}$ & $(-w_y,w_z,-w_x)$			& $\bs{\rho}_{12}$ &$\bs{\rho}_{16}$ & $(-w_y',-w_z',w_x')$		& $\bs{\rho}_{3}$ &$\bs{\rho}_{10}-(1,1,0)$ & $(p_y,-p_z,-p_x)$\\
&$\bs{\rho}_{3}$ & $\bs{\rho}_{16}$ & $(-w_z,w_x,-w_y)$				& $\bs{\rho}_{16}$ &$\bs{\rho}_{5}$ & $(w_z',-w_x',-w_y')$		& $\bs{\rho}_{3}$ &$\bs{\rho}_{11}$ & $(p_z,p_x,p_y)$\\
\hline
&$\bs{\rho}_{4}$ & $\bs{\rho}_{6}$ & $(w_y,-w_z,-w_x)$				& $\bs{\rho}_{6}$ &$\bs{\rho}_{13}$ & $(w_y',w_z',w_x')$			& $\bs{\rho}_{4}$ &$\bs{\rho}_{7}+(0,1,0)$ & $(-p_y,p_z,-p_x)$\\
D'&$\bs{\rho}_{4}$ & $\bs{\rho}_{13}$ & $(w_z,-w_x,-w_y)$			& $\bs{\rho}_{13}$ &$\bs{\rho}_{14}$ & $(-w_z',w_x',-w_y')$		& $\bs{\rho}_{4}$ &$\bs{\rho}_{8}-(1,0,1)$ & $(-p_z,-p_x,p_y)$\\
&$\bs{\rho}_{4}$ & $\bs{\rho}_{14}$ & $(w_x,-w_y,-w_z)$				& $\bs{\rho}_{14}$ &$\bs{\rho}_{6}$ & $(-w_x',-w_y',w_z')$		& ${\cbl\bs{\rho}_{4}}$ &${\cbl\bs{\rho}_{12}}$ & ${\cbl\begin{array}{l}(p_x,p_y,p_z)=\\-(1.4,0.2,0.5)\end{array}}$
\end{tabular}
\end{ruledtabular}
\end{table*}

\end{document}